\documentclass[a4paper,11pt]{article}
\usepackage{jheppub}
\pdfoutput=1

\usepackage[T1]{fontenc}
\usepackage{graphicx,cancel,float}
\usepackage{amssymb}
\usepackage{textcomp}
\usepackage{amsmath}
\usepackage{tabularx}
\usepackage{bm}
\usepackage{times}
\usepackage{color}
\usepackage{slashed}
\usepackage{multirow}
\usepackage{verbatim}
\usepackage{epsfig,epstopdf}

\usepackage{hyperref}
\usepackage{xcolor}
\usepackage{colortbl}
\usepackage{array, boldline, makecell, booktabs}
\usepackage{extarrows}
\usepackage{latexsym}
\definecolor{mygray}{gray}{.95}
\usepackage[title]{appendix}
\allowdisplaybreaks[4]

\def\lsim{\mathrel{\raise.3ex\hbox{$<$\kern-.75em\lower1ex\hbox{$\sim$}}}}
\def\gsim{\mathrel{\raise.3ex\hbox{$>$\kern-.75em\lower1ex\hbox{$\sim$}}}}

\newcommand{\C}{{\tt C}}

\newcommand{\calO}{{\cal O}}

\preprint{}

\title{An explicit construction of the dimension-9 operator basis in the standard model effective field theory}

\author[a,c]{Yi Liao}
\emailAdd{liaoy@nankai.edu.cn}
\author[b]{and Xiao-Dong Ma}
\emailAdd{maxid@phys.ntu.edu.tw}
\affiliation[a]{School of Physics, Nankai University, Tianjin 300071, China}
\affiliation[b]{Department of Physics, National Taiwan University, Taipei 10617, Taiwan}
\affiliation[c]{Center for High Energy Physics, Peking University, Beijing 100871, China}

\abstract
{We investigate systematically dimension-9 operators in the standard model effective field theory which contains only standard model fields and respects its gauge symmetry. With the help of the Hilbert series approach to classifying operators according to their lepton and baryon numbers and their field contents, we construct the basis of operators explicitly. We remove redundant operators by employing various kinematic and algebraic relations including integration by parts, equations of motion, Schouten identities, Dirac matrix and Fierz identities, and Bianchi identities. We confirm counting of independent operators by analyzing their flavor symmetry relations. All operators violate lepton or baryon number or both, and are thus non-Hermitian. Including Hermitian conjugated operators there are $384|^{\Delta L=\pm 2}_{\Delta B=0}+10|^{\Delta L=0}_{\Delta B=\pm 2}+4|^{\Delta L=\pm3}_{\Delta B=\pm1}+236|^{\Delta L=\mp 1}_{\Delta B=\pm1}$ operators without referring to fermion generations, and $44874|^{\Delta L=\pm 2}_{\Delta B=0}+2862|^{\Delta L=0}_{\Delta B=\pm 2}+486|^{\Delta L=\pm3}_{\Delta B=\pm1}+42234|^{\Delta L=\pm 1}_{\Delta B=\mp1}$ operators when three generations of fermions are referred to, where $\Delta L,~\Delta B$ denote the net lepton and baryon numbers of the operators. Our result provides a starting point for consistent phenomenological studies associated with dimension-9 operators.
}


\begin{document}
\maketitle
\flushbottom


\setcounter{page}{1}

\section{Introduction}
\label{sec1}

The standard model has been verified to be a good effective field theory that works successfully below the electroweak scale $\Lambda_{\rm EW}$ of a few hundreds GeV. The null result in searching for new particles implies that new physics which presumably holds the key to tiny neutrino masses and fundamental issues such as origin of spontaneous symmetry breaking should exist at a scale $\Lambda_{\rm NP}$ significantly higher than $\Lambda_{\rm EW}$. In this circumstance the new physics effects in currently available experiments may be comfortably described by an effective field theory defined between the above two scales. This effective field theory, termed the standard model effective field theory (SMEFT), contains exclusively the dynamical degrees of freedom in the standard model (SM) and respects its gauge symmetry $SU(3)_C\times SU(2)_L\times U(1)_Y$. While the SM interactions are dominant, effective interactions originating from new physics are important and interesting -- they could modify SM predictions in precision measurements or induce new physical effects that break accidental symmetries in SM such as baryon and lepton number conservation. They therefore deserve a persistent and systematic investigation.

The effective interactions appear as higher-dimensional operators with effective couplings called Wilson coefficients. While the higher-dimensional operators in SMEFT are determined in terms of the SM fields and gauge symmetry, their Wilson coefficients are completely unknown and encode the information of new physics lying at a high scale $\Lambda_{\rm NP}$. To bridge the measurements at low energies to new physics at a high scale, an important task is to establish a basis of operators at each dimension and analyze their renormalization group evolution between the two scales. While the unique dimension-5 (dim-5) operator~\cite{Weinberg:1979sa} has been known for a long time, the bases for even higher dimensional operators have only been established in recent years.  At dimension 6, there are 63 independent operators without counting fermion flavors~\cite{Abbott:1980zj,Buchmuller:1985jz,Grzadkowski:2010es}, among which 59 conserve both lepton and baryon numbers and the other 4 violate both by one unit ($\Delta L=\Delta B=\pm1$).  At dimension 7 it turns out that there are 12 operators that violate lepton number by two units ($\Delta L=\pm 2$) and 6 operators that violate both lepton and baryon numbers by one unit ($-\Delta L=\Delta B=\pm1$)~\cite{Lehman:2014jma,Liao:2016hru}. For the first time the flavor relations among effective operators start to become nontrivial making counting of independent operators complicated. As the dimension increases further, even counting of complete and independent operators becomes difficult. Fortunately, there is an approach based on the Hilbert series (HS)~\cite{Jenkins:2009dy,Hanany:2010vu} that can be used to enumerate the total number of independent operators for a given configuration of fields at each dimension~\cite{Lehman:2015via, Lehman:2015coa,Henning:2015alf,Henning:2017fpj}; automatic tools based on the HS have been developed~\cite{Henning:2015alf}, see Refs.~\cite{Gripaios:2018zrz,Criado:2019ugp, Fonseca:2019yya, Marinissen:2020jmb,Banerjee:2020bym} for recent efforts. Although the HS does not tell the concrete forms of operators, it is very helpful for us to exhaust all possible forms and confirm their independency as we saw in the construction of dim-7 operators~\cite{Liao:2016hru} and very recently of dim-8 operators~\cite{Li:2020gnx,Murphy:2020rsh}. If there are new light particles beyond the SM that are most likely a SM singlet such as light sterile neutrinos, they must be incorporated into the framework of effective field theory thus extending the regime of the SMEFT~\cite{Aparici:2009fh,delAguila:2008ir,Bhattacharya:2015vja,Liao:2016qyd}.

Our goal in this work is to push this endeavor one step further by constructing a basis of dim-9 operators in SMEFT.  Some small subsets of dim-9 operators have previously appeared in the literature focusing on processes of specific interest, including the operators related to Majorana neutrino mass and nuclear neutrinoless double beta decays ($0\nu\beta\beta$)~\cite{Babu:2001ex,deGouvea:2007qla,Graesser:2016bpz,Gustafsson:2020bou}, the operators contributing to neutron-antineutron ($n-\bar n$) oscillation~\cite{Rao:1983sd,Caswell:1982qs,Rinaldi:2018osy}, and those relevant to rare nucleon decays~\cite{Weinberg:1980bf,Weldon:1980gi,Hambye:2017qix}. This work provides the first systematic investigation on the basis of complete and independent dim-9 operators, and would serve as the starting point for further physical analysis.

This paper is organized as follows. In section~\ref{sec2}, we will set up our notations and conventions, and summarize various identities to be used for the reduction of redundant operators. In section~\ref{sec3}, we will establish the basis for dim-9 operators in SMEFT. We will illustrate class by class as shown in table~\ref{tab1} how to perform the reduction to reach the final basis given in tables~\ref{tab:LV1}-\ref{tab:BV3}. We summarize our main results and mention very briefly possible phenomenology related to dim-9 operators in section~\ref{sec4}. For operators with flavor symmetries we summarize their flavor relations in appendix~\ref{app:flavorsymmetry}, which have helped us count operators independently of the HS approach and thus confirm our basis.

\section{Preliminaries}
\label{sec2}

We start with some notations and conventions. The SMEFT is an effective field theory above the electroweak scale $\Lambda_{\rm EW}$ but far below some new physics scale $\Lambda_{\rm NP}$. It inherits the SM gauge group $SU(3)_C\otimes SU(2)_L\otimes U(1)_Y$ and field contents: the $n_f(=3)$ generations of left-handed lepton and quark doublet fields $L(1,2,-1/2)$, $Q(3,2,1/6)$ and right-handed up-type quark, down-type quark and charged lepton singlet fields $u(3,1,2/3)$, $d(3,1,-1/3)$, $e(1,1,-1)$, and the Higgs doublet $H(1,2,1/2)$. Its effective Lagrangian contains the SM Lagrangian as the leading terms that is augmented by effective interactions involving operators of higher and higher dimensions. To set up our notations and conventions, we write down the SM Lagrangian:
\begin{eqnarray}
\label{sml}
\nonumber
\mathcal{L}_{\rm SM}&=&
-\frac{1}{4}G^A_{\mu\nu}G^{A\mu\nu}-\frac{1}{4}W^I_{\mu\nu}W^{I\mu\nu} -\frac{1}{4}B_{\mu\nu}B^{\mu\nu}
\\
&&+(D_\mu H)^\dagger(D^\mu H)
-\lambda\Big(H^\dagger H-\frac{1}{2}v^2\Big)^2
\nonumber
\\
&&+\sum_{\psi=Q, L, u, d, e}\overline{\psi}i \slashed{D}\psi
-\left[\overline{L}Y_e e H+\overline{Q}Y_u u \tilde{H}+\overline{Q}Y_d d H +\rm{h.c.}\right]\;.
\end{eqnarray}
Here the superscripts $A$ and $I$ count the generators of the gauge groups $SU(3)_C$ and $SU(2)_L$ respectively, $Y_u,~Y_d,~Y_e$ are the Yukawa couplings which are complex matrices in the flavor space, and $\tilde H=i\tau_2 H^*$. For $n_f$ generations of fermions, we label the fermion fields by the subscript Latin letters $(p,r,s,t,v,w)$; for instance, $L_p$ is the $p$-th generation left-handed lepton doublet. The covariant derivative is defined by
\begin{eqnarray}
D_\mu=\partial_\mu-ig_3 T^AG^A_\mu-ig_2T^IW^I_\mu-ig_1YB_\mu,
\end{eqnarray}
where $g_{1,2,3}$ are the gauge couplings, and $Y,~T^I,~T^A$ are the generators appropriate for the fields to be acted on. We use  the superscript Latin letters ($i,j,k,l,m,n$) and Greek letters ($\alpha,\beta,\gamma,\rho,\sigma,\tau$) to denote field components in the fundamental representations of $SU(2)_L$ and $SU(3)_C$ respectively. We define the following shortcuts for better presentation of operators:
\begin{align}
\nonumber
W^{\mu\nu}\equiv& W^{I\mu\nu}\tau^I\;,~ &
(\epsilon W^{\mu\nu})_{ij}\equiv& \epsilon_{ik}W^{\mu\nu}_{kj}= \epsilon_{ik}(\tau^I)_{kj} W^{I\mu\nu}\;,~
\\
G^{\mu\nu}\equiv& G^{A\mu\nu}\lambda^A\;,~ &
(\epsilon G^{\mu\nu})_{\alpha\beta\gamma}\equiv&\epsilon_{\alpha\beta\tau}
G^{\mu\nu}_{\tau\gamma}
=\epsilon_{\alpha\beta\tau}(\lambda^A)_{\tau \gamma}G^{A\mu\nu}\;,
\end{align}
where $\tau^I$ and $\lambda^A$ are the Pauli and Gell-Mann matrices, respectively. The dual field tensor is defined to be $\tilde X^{\mu\nu}=\epsilon^{\mu\nu\rho\sigma}X_{\rho\sigma}/2$ with $X=B,~W,~G$.

We will systematically classify and construct all dim-9 operators in the next section. To construct a basis for the operators, we first generate all possible field configurations from the HS~\cite{Henning:2015alf}, i.e., all subsets of ingredients (fermion and scalar fields, covariant derivative, and gauge field strength tensors) including the number of each ingredient that together can form a gauge and Lorentz invariant dim-9 operator. Then, for each field configuration we construct explicitly complete and independent operators whose total number is consistent with the HS counting. This counting of operators is further confirmed independently by employing flavor relations described in appendix~\ref{app:flavorsymmetry}. The construction of operators and removal of all redundant ones is highly nontrivial because with so many ingredients which have rich Lorentz and gauge group properties one could make many apparently different but actually related contractions in Lorentz and gauge group indices. We will utilize all kinematical and algebraic weapons in our arsenal to reach the goal, which include integration by parts relations (IBP), the SM equations of motion (EoM), various Schouten identities (SI) for totally antisymmetric rank-$n$ tensors (with $n=2,3,4$), Dirac gamma matrix identities (GI), generalized Fierz identities (FI), and the Bianchi identity (BI) for gauge fields. In the rest of this section we summarize these relations and identities.

Let us start with Schouten identities which are useful in relating various Lorentz and gauge contractions involving totally antisymmetric constant tensors that would look independent. For the $SU(2)$ group, we have the SIs involving the rank-2 tensor $\epsilon_{ij}$,
\begin{align}
&\delta_{ij}\epsilon_{kl}+\delta_{ik}\epsilon_{lj}+\delta_{il}\epsilon_{jk}=0\;, ~~
&&\epsilon_{ij}\epsilon_{kl}+\epsilon_{ik}\epsilon_{lj}+\epsilon_{il}\epsilon_{jk}
=0\;.
\label{SI:r2}
\end{align}
For the $SU(3)$ group, we obtain the following SIs involving the rank-3 tensor $\epsilon_{\alpha\beta\gamma}$,
\begin{align}
\nonumber
&\delta_{\alpha\beta}\epsilon_{\rho\sigma\tau}
-\delta_{\alpha\rho}\epsilon_{\sigma\tau\beta}
+\delta_{\alpha\sigma}\epsilon_{\tau\beta\rho}
-\delta_{\alpha\tau}\epsilon_{\beta\rho\sigma}=0\;,
\\
\nonumber
&\epsilon_{\alpha\beta\gamma}\epsilon_{\rho\sigma\tau}
-\epsilon_{\alpha\beta\rho}\epsilon_{\sigma\tau\gamma}
+\epsilon_{\alpha\beta\sigma}\epsilon_{\tau\gamma\rho}
-\epsilon_{\alpha\beta\tau}\epsilon_{\gamma\rho\sigma}=0\;,
\\
&(\epsilon_{\alpha\beta\gamma}\epsilon_{\rho\sigma\tau}
+\epsilon_{\alpha\beta\sigma}\epsilon_{\rho\tau\gamma}
+\epsilon_{\alpha\beta\tau}\epsilon_{\rho\gamma\sigma})
+(\epsilon_{\alpha\sigma\tau}\epsilon_{\rho\beta\gamma}
+\epsilon_{\alpha\tau\gamma}\epsilon_{\rho\beta\sigma}
+\epsilon_{\alpha\gamma\sigma}\epsilon_{\rho\beta\tau})=0\;.
\label{SI:r3}
\end{align}
For the Lorentz group we will need the following SI involving the rank-4 tensor $\epsilon_{\rho\sigma\tau\eta}$,
\begin{align}
g_{\mu\nu}\epsilon_{\rho\sigma\tau\eta}
+g_{\mu\rho}\epsilon_{\sigma\tau\eta\nu}
+g_{\mu\sigma}\epsilon_{\tau\eta\nu\rho}
+g_{\mu\tau}\epsilon_{\eta\nu\rho\sigma}
+g_{\mu\eta}\epsilon_{\nu\rho\sigma\tau}=0\;.
\label{SI:r4}
\end{align}
Besides the above standard SIs, we will also require identities that involve  generators in the fundamental representations of $SU(2)$ and $SU(3)$:
\begin{align}
&\epsilon_{ik}(\tau^I)_{kj}=\epsilon_{jk}(\tau^I)_{ki}\;,~
&\epsilon_{\alpha\beta\tau}(\lambda^A)_{\tau\gamma}
+\epsilon_{\beta\gamma\tau}(\lambda^A)_{\tau\alpha}
+\epsilon_{\gamma\alpha\tau}(\lambda^A)_{\tau\beta}
=0\;,
\end{align}
which imply that
\begin{align}
&(\epsilon W^{\mu\nu})_{ij}=(\epsilon W^{\mu\nu})_{ji},
\nonumber
\\
&(\epsilon G^{\mu\nu})_{\alpha\beta\gamma}
+(\epsilon G^{\mu\nu})_{\beta\gamma\alpha}
+(\epsilon G^{\mu\nu})_{\gamma\alpha\beta}=0.
\end{align}

Now we collect some identities for Dirac gamma matrices which will be used to simplify fermion structures:
\begin{align}
\nonumber
&\sigma^{\mu\nu}=i\gamma^\mu\gamma^\nu-ig^{\mu\nu}\;,
\nonumber
\\
&\epsilon^{\mu\nu\rho\sigma}\gamma_\sigma P_\pm
=\pm i(\gamma^\mu\gamma^\rho\gamma^\nu-g^{\mu\rho}\gamma^\nu
-g^{\rho\nu}\gamma^\mu+g^{\mu\nu}\gamma^\rho)P_\pm\;,
\nonumber
\\
&\epsilon^{\mu\nu\rho\sigma}\sigma_{\rho\sigma}P_\pm
=\mp 2i\sigma^{\mu\nu}P_\pm\;,
\nonumber
\\
&\epsilon^{\mu\nu\rho\eta}\sigma^\sigma_{~\eta} P_\pm
=\pm i(g^{\nu\sigma}\sigma^{\mu\rho}-g^{\mu\sigma}\sigma^{\nu\rho}
-g^{\rho\sigma}\sigma^{\mu\nu})P_\pm\;,
\label{GI}
\end{align}
with the chiral projectors $P_\pm=(1\pm\gamma_5)/2$. The above are straightforward to derive; see, for instance Ref.~\cite{Liao:2012uj}. These seemingly trivial identities will be applied judiciously together with the Fierz identities that we now describe. The Fierz identities are essentially identities for direct products of Dirac matrices, and become identities for products of fermion bilinears when combined with Dirac spinors. We will sometimes denote $(\overline{\psi_1}\Gamma_1\psi_2)(\overline{\psi_3}\Gamma_2\psi_4)$ as $\Gamma_1\otimes\Gamma_2$ and its Fierz transformed partner $(\overline{\psi_1}\Gamma_1\psi_4)(\overline{\psi_3}\Gamma_2\psi_2)$ as $\Gamma_1\odot\Gamma_2$. In applying Fierz identities for operators it is necessary to include a minus sign when interchanging the order of two fermion fields.

Our first set of Fierz identities are the standard completely contracted products of two bilinears:
\begin{align}\nonumber
&( \overline{\psi_{1}} \gamma^\mu P_\pm \psi_{2})( \overline{\psi_{3}}  \gamma_\mu P_\mp\psi_{4})=-2(\overline{\psi_{1}}P_\mp \psi_{4})(\overline{\psi_{3}}P_\pm\psi_{2})\;,
\\\nonumber
&( \overline{\psi_{1}} \gamma^\mu P_\pm \psi_{2})( \overline{\psi_{3}}  \gamma_\mu P_\pm \psi_{4})=+2(\overline{\psi_{1}}P_\mp \psi^\C_{3})(\overline{\psi^\C_{4}}P_\pm\psi_{2})\;,
\\\nonumber
&( \overline{\psi_{1}} \sigma^{\mu\nu} P_{\pm}\psi_{2})( \overline{\psi_{3}} \sigma_{\mu\nu} P_{\mp}\psi_{4})=0\;,
\\
&( \overline{\psi_1}\sigma_{\mu\nu} P_{\pm}\psi_2)(\overline{\psi_3}\sigma^{\mu\nu} P_{\pm}\psi_4)=-4(\overline{\psi_1} P_{\pm}\psi_2)(\overline{\psi_3} P_{\pm}\psi_4)-8(\overline{\psi_1} P_{\pm}\psi_4)(\overline{\psi_3} P_{\pm} \psi_2)\;.
\label{FI1}
\end{align}
The point here is that contracted vector and tensor products are expressed exclusively in terms of products of scalar bilinears. For charge conjugation of chiral fields, we denote $\psi^\C_\pm\equiv(\psi_\pm)^\C=C\overline{\psi_\pm}^{\rm T}$, where the charge conjugation matrix $C$ satisfies $C^{\rm T}=C^\dagger=-C$ and $C^2=-1$ so that $\psi_\pm=(\psi^\C_\pm)^\C$. And for brevity we also denote $D_\mu \psi^{\C,i\alpha}_p=(D_\mu \psi_p)^{\C,i\alpha}$ with the superscript $x$ being the weak isospin index $i$ and/or color index $\alpha$ and the subscript $p$ the generation index. Our second set of FIs are,
\begin{align}
\nonumber
&( \overline{\psi_1} P_\pm \psi_2)( \overline{\psi_3}P_\pm\psi_4)=-( \overline{\psi_1}P_\pm\psi_{3}^\C)( \overline{\psi^\C_2}P_\pm\psi_4)-( \overline{\psi_1}P_\pm\psi_4)( \overline{\psi_3}P_\pm \psi_2)\;,
\\
\nonumber
&( \overline{\psi_1}\gamma^\mu P_\pm \psi_2)( \overline{\psi_3}P_\pm\psi_4)=-( \overline{\psi_1}\gamma^\mu P_\pm\psi_{3}^\C)( \overline{\psi^\C_2}P_\pm\psi_4)-( \overline{\psi_1}\gamma^\mu P_\pm\psi_4)( \overline{\psi_3}P_\pm \psi_2)\;,
\\
&( \overline{\psi_1}\sigma^{\mu\nu} P_\pm \psi_2)( \overline{\psi_3}P_\pm\psi_4)=-( \overline{\psi_1}\sigma^{\mu\nu}P_\pm\psi_{3}^\C)( \overline{\psi^\C_2}P_\pm\psi_4)-( \overline{\psi_1}\sigma^{\mu\nu}P_\pm\psi_4)( \overline{\psi_3}P_\pm \psi_2)\;.
\label{FI2}
\end{align}
One can see the second and third ones can be obtained from the first by replacing $\overline{\psi_1}$ by $\overline{\psi_1}\gamma^\mu$ and $\overline{\psi_1}\sigma^{\mu\nu}$, respectively. Such a replacement is also valid for any other field in equations~\eqref{FI1} and \eqref{FI2}, including the case when $\psi_i$ is a charge conjugated field $\psi^\C$.

Finally, we will need the Bianchi identity for gauge fields. For a field strength tensor $X_{\mu\nu}$, the BI means
\begin{align}
D_\mu X_{\nu\rho}+D_\nu X_{\rho\mu}+D_\rho X_{\mu\nu}=0\;,~~{\rm or}~~
D_\nu \tilde X^{\mu\nu}=0\;.
\label{BI}
\end{align}
This will be useful in reducing operators containing both field strength tensors and covariant derivatives.

\section{Operator reduction and operator basis at dimension nine in the SMEFT}
\label{sec3}

\begin{table}
\centering
\resizebox{\linewidth}{!}{
\renewcommand{\arraystretch}{1}
\begin{tabular}{| c | c | c | c | c | }
\hline
Sectors& $(\Delta B,\Delta L)=(0,2)$  & $(\Delta B,\Delta L)=(1,3)$ & $(\Delta B,\Delta L)=(2,0)$ &  $(\Delta B,\Delta L)=(1,-1)$
\\\hline
\# for $n_f=3$  & 21117 & 243 & 1431 & 22437
\\\hline
\multirow{3}{*}{Classes}
&\multicolumn{4}{|c|}{$\psi^2\varphi^6,\quad \psi^2\varphi^4X,\quad \psi^2\varphi^2X^2,\quad \psi^4\varphi X(\checkmark),\quad  \psi^4\varphi^3(\checkmark),\quad \psi^6$\color{red}(\checkmark)}
\\
& \multicolumn{4}{|c|}{$ \psi^2\varphi^5D,\quad    \psi^2\varphi^3XD,\quad  \psi^4\varphi^2D(\checkmark),\quad   \psi^4XD(\checkmark)$ }
\\
& \multicolumn{4}{|c|}{$ \psi^2\varphi^4D^2,\quad \psi^2\varphi^2XD^2,\quad \psi^4\varphi D^2(\checkmark),\quad \psi^2\varphi^3D^3,\quad \psi^4D^3(\checkmark) ,\quad \psi^2\varphi^2D^4  $ }
\\\hline
\end{tabular}
}
\caption{Dim-9 operators fall into various sectors and classes according to their lepton and baryon numbers and their field configurations, with the number of independent operators shown for each sector with three generations of fermions. Hermitian conjugates of operators are not included.}
\label{tab1}
\end{table}

To begin with, we first run the {\em Mathematica} code for the HS method~\cite{Henning:2015alf} to obtain the field configurations of dim-9 operators and to count the corresponding total number of independent operators for each field configuration satisfying gauge and Lorentz invariance. Since dim-9 operators necessarily violate baryon or lepton number or both in various ways~\cite{Kobach:2016ami}, they are non-Hermitian. These operators thus naturally fall into four sectors according to the baryon and lepton numbers they carry, without counting hermitian conjugates. We will discuss and list one half of them, and the other half is obtained by Hermitian conjugation. The operators can also be classified by their field configurations. We will use a generic fermion field $\psi$, scalar field $\varphi$ and field strength tensor $X$ to denote the SM fermions $\{Q,L,u,d,e\}$ and their conjugates, the Higgs $\{H, H^*\}$ and gauge fields $\{B,W,G\}$, respectively. These sectors and classes are displayed in table~\ref{tab1}, together with a count of independent operators in each sector for three generations of fermions. A check mark following a class indicates that it contains baryon-number-violating operators. Our final basis for dim-9 operators is summarised in tables~\ref{tab:LV1}, \ref{tab:LV2}, \ref{tab:LV3}, \ref{tab:LV4} and \ref{tab:LV5} for the sector $(\Delta B,\Delta L)=(0,2)$,  the upper part of table~\ref{tab:BV1} for the sectors $(\Delta B,\Delta L)=(1,3),~(2,0)$, and the lower part of table~\ref{tab:BV1}, tables~\ref{tab:BV2} and \ref{tab:BV3} for the sector $(\Delta B,\Delta L)=(1,-1)$, respectively. In all tables, we also count the number of independent operators for $n_f$ generations of fermions, which helps to confirm our results against the HS approach. We stress again that the complete basis of dim-9 operators includes the Hermitian conjugates of all listed operators.

Here we make some comments on conventions. (1) Each gauge field strength is associated with its gauge coupling, i.e., $\{g_1 B, g_2 W, g_3G\}$. (2) An imaginary unit $i$ is attached to gauge covariant derivative $D$, field strength tensor $X$, and Dirac tensor $\sigma_{\mu\nu}$, in order to avoid some superficial $i$ factors in the calculation of renormalization group effects and $S$-matrix elements. (3) Some obvious contractions of $SU(2)_L$ and $SU(3)_C$ indices by the Kronecker deltas $\delta_{ij}$ and $\delta_{\alpha\beta}$ are suppressed for brevity. In the following subsections we will illustrate our construction steps and techniques of the operator basis class by class. Our main strategy is that for field configurations involving four or six fermions we first employ FIs to fix the fermion structures, then finish $SU(2)_L$ and $SU(3)_C$ contractions, and finally exhaust insertions of covariant derivatives.

\subsection{Classes $\psi^2\varphi^6$, $\psi^2\varphi^4X$, and $\psi^2\varphi^2X^2$}

The class $\psi^2\varphi^6$ has a single field configuration $L^2H^4H^{*2}$ which leads to the unique dim-9 neutrino mass operator shown in table~\ref{tab:LV1}. Such Majorana neutrino mass operators have been systematically classified at any dimension in~\cite{Liao:2010ku,Liao:2017amb}. The class $\psi^2\varphi^4X$ also has a single field configuration $L^2H^3H^*X$ with field strength $X=B,~W$. The operators are straightforward generalizations of similar dim-7 operators in the class $\psi^2\varphi^2X$ by a Higgs pair $H^\dagger H$~\cite{Lehman:2014jma,Liao:2016hru}.  They are also shown in table~\ref{tab:LV1}.

The SM gauge symmetry restricts the field configuration in the class $\psi^2\varphi^2X^2$ to be $L^2H^2X^2$ with $X=B,~W,~G$. The operators follow three types of Lorentz contractions:
\begin{align}
&(LL)HHX_1^{\mu\nu}X_{2\mu\nu}\;,
&&(LL)HHX_1^{\mu\nu}\tilde X_{2\mu\nu}\;,
 &&(L\sigma_{\mu\nu}L)HHX_1^{\mu\rho}X^{\nu}_{2~\rho}\;.
\end{align}
Note that the tensor type with a dual field strength tensor is not independent by the last GI in equation~\eqref{GI}. In addition, the tensor operator vanishes for $X_1=X_2$ by symmetry arguments, but can survive with $(X_1,X_2)=(B,W)$. The final results are shown in table~\ref{tab:LV1}. For the case with  $X_1=X_2=W$, the following identities are used to reduce redundant operators due to different $SU(2)_L$ contractions,
\begin{align}
\nonumber
&(\epsilon W^{\mu\nu})_{ij}(\epsilon W^{\mu\nu})_{kl}=(\epsilon W^{\mu\nu})_{ik}(\epsilon W^{\mu\nu})_{jl}
-\epsilon_{il}\epsilon_{jk}W^I_{\mu\nu} W^{I,\mu\nu}\;,
\\
&(\epsilon W^{\mu\nu})_{ij}(\epsilon \tilde W^{\mu\nu})_{kl}=(\epsilon W^{\mu\nu})_{ik}(\epsilon \tilde W^{\mu\nu})_{jl}
-\epsilon_{il}\epsilon_{jk}W^I_{\mu\nu}\tilde W^{I,\mu\nu}\;.
\end{align}
Their proofs are straightforward using the second SI in equation~\eqref{SI:r2}. The operators in the class $\psi^2\varphi^2X^2$ can contribute directly to the Majorana neutrino-photon scattering which has been studied in Ref.~\cite{Liao:2020zyx}.

\subsection{Classes $\psi^4\varphi X$ and $\psi^4\varphi^3$}

The operators in the class $\psi^4\varphi X$ are first classified according to their fermion structures which are a product of two bilinears. Since the field strength tensor has two Lorentz indices, we have three possible structures: the scalar-tensor ($ST$), vector-vector ($VV$), and tensor-tensor ($TT$) ones, which can be generically parameterized as follows:
\begin{align}
\nonumber
&{\cal O}_{ST}:~(\overline{\psi_1}\psi_2)
(\overline{\psi_3}\sigma_{\mu\nu}\psi_4)X^{\mu\nu}\varphi\;,
\\
&{\cal O}_{VV}:~(\overline{\psi_1}\gamma_\mu \psi_2)
(\overline{\psi_3}\gamma_\nu\psi_4)X^{\mu\nu}\varphi\;,~
\nonumber
\\
&{\cal O}_{TT}:~(\overline{\psi_1}\sigma_{\mu\rho} \psi_2)
(\overline{\psi_3}\sigma^{\rho}_{~\nu}\psi_4)X^{\mu\nu}\varphi\;.
\end{align}
However, we find the last two can be transformed into the first one by the FIs:
\begin{align}
\nonumber
&(2\gamma_\mu P_\pm  \otimes \gamma_\nu P_\mp)X^{\mu\nu}
=i\left(\sigma^{\mu\nu}P_\mp\odot P_\pm-P_\mp\odot \sigma^{\mu\nu} P_\pm  \right)X^{\mu\nu}\;,
\\
&(\sigma_{\mu\rho}P_\pm \otimes \sigma^{~\rho}_{\nu}P_\mp)X^{\mu\nu}=0\;,~
\nonumber
\\
&(\sigma_{\mu\rho}P_\pm \otimes \sigma^{~\rho}_{\nu}P_\pm)X^{\mu\nu}
=i\left(\sigma_{\mu\nu}P_\pm \odot P_\pm-P_\pm \odot \sigma_{\mu\nu} P_\pm  \right)X^{\mu\nu}\;.
\label{FI3}
\end{align}
Therefore, we only need to consider ${\cal O}_{ST}$. Note in addition that ${\cal O}_{ST}$ with a dual tensor $\tilde X$ in place is equivalent to ${\cal O}_{ST}$ due to the third GI in equation~\eqref{GI}. Taking the SM field contents into consideration and from the HS output we can write down independent operators in each field configuration that are allowed by the SM gauge symmetry. For some field configurations with multiple $SU(2)_L$ doublets there are many apparently independent ways to perform $SU(2)_L$ contractions which would yield redundant operators. Those redundant operators can be reduced by the SIs in equation~\eqref{SI:r2} and the following derived identities involving the $SU(2)_L$ field strength tensor $W$,
\begin{align}
&\epsilon_{ij}(\epsilon W^{\mu\nu})_{kl}=\epsilon_{il}(\epsilon W^{\mu\nu})_{jk}-\epsilon_{jl}(\epsilon W^{\mu\nu})_{ik}\;,~
\nonumber
\\
&\delta_{ij}(\epsilon W^{\mu\nu})_{kl}=\delta_{ik}(\epsilon W^{\mu\nu})_{jl}-(W^{\mu\nu})_{il}\epsilon_{jk}\;.
\label{SIW}
\end{align}
In addition, FIs in equation~\eqref{FI2} and its generalizations (e.g., by replacing $\psi_2$ by $\sigma_{\mu\nu}\psi_2$ in the first identity in equation~\eqref{FI2}) are also required to reduce redundant operators when identical fermions are present. The final results are listed in table~\ref{tab:LV1} for the $(\Delta B, \Delta L)=(0,2)$ sector and table~\ref{tab:BV1} for the $(\Delta B, \Delta L)=(1,-1)$ sector. As an example of the above manipulations, we consider the following operator from the field configuration $Qd^*L^2HB$:
\begin{align}
&~~~~~~g_1\epsilon_{ik}\epsilon_{jl}(\overline{d_p}\sigma_{\mu\nu}  L^i_r)
(\overline{Q^{\C,j}_s}L^k_t)H^lB^{\mu\nu}
\nonumber
\\
&\overset{\rm SI}{=}
+g_1\epsilon_{il}\epsilon_{jk}(\overline{d_p}\sigma_{\mu\nu}  L^i_r)
(\overline{Q^{\C,j}_s}L^k_t)H^lB^{\mu\nu}+\calO_{dLQLHB3}^{prst}
\nonumber
\\
&\overset{\rm FI}{=}-g_1\epsilon_{il}\epsilon_{jk}
(\overline{d_p}\sigma_{\mu\nu}  Q^j_s)(\overline{L^{\C,i}_r}L^k_t)H^lB^{\mu\nu}
+\calO_{dLQLHB3}^{prst}+\calO_{dLQLHB3}^{ptsr}
\nonumber
\\
&\overset{\rm FI}{=}
-g_1\epsilon_{il}\epsilon_{jk}
(\overline{d_p}L^i_r)(\overline{Q^{\C,j}_s}\sigma_{\mu\nu}L^k_t)H^lB^{\mu\nu}
+\calO_{dLQLHB3}^{prst}+\calO_{dLQLHB3}^{ptsr}+\calO_{dLQLHB1}^{ptsr}
\nonumber
\\
&\overset{\rm SI}{=}
+\calO_{dLQLHB3}^{prst}+\calO_{dLQLHB3}^{ptsr}+\calO_{dLQLHB1}^{prst}
+\calO_{dLQLHB1}^{ptsr}-\calO_{dLQLHB2}^{prst}\;,
\end{align}
where the first FI refers to the third FI in equation~\eqref{FI2} and the second FI is obtained by treating $\sigma_{\mu\nu}Q^j_s$ as $\psi_2$ in the first FI in equation~\eqref{FI2}.

The four-fermion part in the class $\psi^4\varphi^3$ can always be written as a product of two scalar bilinears by FIs summarised in equation~\eqref{FI1}. Taking the SM field contents into consideration and from the HS output, we can readily write down the operators for each allowed field configuration. The final results are shown in table~\ref{tab:LV2} for the $(\Delta B, \Delta L)=(0,2)$ sector and table~\ref{tab:BV1} for the $(\Delta B, \Delta L)=(1,-1)$ sector, respectively. In obtaining the basis, SIs in equation~\eqref{SI:r2} should be repeatedly implemented to reduce redundant operators arising from various $SU(2)_L$ contractions among multiple $SU(2)_L$ doublets. One can show that any other operators in this class can be expressed as linear combinations of the listed operators with the aid of FIs in equations~\eqref{FI1}-\eqref{FI2} and SIs in equation~\eqref{SI:r2}.

\subsection{Class $\psi^6$}
\label{secpsi6}
We first show that all operators in this class can be written as a product of three scalar fermion bilinears ($SSS$). This is realized by the following FIs applied to three bilinears:
\begin{align}\nonumber
(\sigma^{\mu\rho} P_{\pm}\otimes\sigma^{~~\nu}_{\rho}P_{\mp})
\otimes \sigma_{\mu\nu}P_{\pm} =&0,
\\
\nonumber
(\sigma^{\mu\rho} P_{\pm}\otimes\sigma^{~~\nu}_{\rho}P_{\pm})
\otimes \sigma_{\mu\nu}P_{\pm} =&
i\left(\sigma^{\mu\nu}P_\pm\odot P_\pm-P_\pm\odot \sigma^{\mu\nu} P_\pm  \right)\otimes \sigma_{\mu\nu}P_{\pm},
\\
\nonumber
2(\gamma^\mu P_{\pm}\otimes\gamma^\nu P_{\mp})
\otimes \sigma_{\mu\nu}P_{\pm}=&
i\left(\sigma^{\mu\nu}P_\mp\odot P_\pm-P_\mp\odot \sigma^{\mu\nu} P_\pm  \right)\otimes \sigma_{\mu\nu}P_{\pm},
\\
\nonumber
\gamma_\mu P_{\pm}\otimes(\gamma_\nu P_{\pm}\otimes \sigma^{\mu\nu}P_{\pm})
=&i\gamma_\mu P_{\pm}\otimes
\left(\gamma^\mu P_\pm \otimes P_\pm +2\gamma^\mu P_\pm \odot P_\pm  \right)\;,
\\
\gamma_\mu P_{\pm}\otimes(\gamma_\nu P_{\pm}\otimes \sigma^{\mu\nu}P_{\mp})
=&-i \gamma_\mu P_{\pm}\otimes\left(\gamma^\mu P_\pm \otimes P_\mp +2 P_\mp \odot \gamma^\mu P_\pm  \right)\;.
\label{FI4}
\end{align}
The terms on the right-hand side only involve $VVS$ or $TTS$ bilinears which can be further transformed into $SSS$ structures by FIs in equation~\eqref{FI1}.  We therefore conclude that all possible dim-9 six-fermion operators can be written as a pure $SSS$ bilinear form. Nevertheless, in the sector $(\Delta B, \Delta L)=(0,2)$ for operators involving four quarks and two leptons, we do not fully follow this convention, but prefer to parameterize operators as a quark-lepton separated form in order to compare easily with the results in the literature (see table~\ref{tab:LV3}). For other sectors with baryon number violation we parameterize operators in the pure scalar form as shown in table~\ref{tab:BV1} for the $(\Delta B, \Delta L)=(2,0)$ and $(1,3)$ sectors and table~\ref{tab:BV2} for the $(\Delta B, \Delta L)=(1,-1)$ sector (except for the two operators ${\cal O}_{LLeudd}$ and ${\cal O}_{LQdddu}$ which have an $STT$ structure). Due to the complexity of operators in the $\psi^6$ class, we analyze each sector separately to confirm our results.

\noindent
{$\scriptscriptstyle\blacksquare$ $(\Delta B, \Delta L)=(0,2)$  sector}

From the HS output the field configurations in this sector can be further divided into three cases: operators with six leptons, operators with two quarks and four leptons, and operators with four quarks and two leptons, which are separated by thick lines in table~\ref{tab:LV2}. For operators with six leptons the only field configuration is $L^4ee^*$ which leads to the unique scalar-type operator shown in table~\ref{tab:LV2}. For operators with two quarks and four leptons, there are four field configurations all of which are written as a scalar. Finally, for operators with four quarks and two leptons, we parameterize  them as a quark-lepton separated form except for the configuration $u^2d^{*2}e^2$. The completeness and independence of the operators for each allowed field configuration is guaranteed by FIs in equations~\eqref{FI1}, \eqref{FI2}, \eqref{FI4}, \eqref{FI5} and SIs in equation~\eqref{SI:r2}. In Ref.~\cite{Babu:2001ex}, Babu and Leung listed 12 six-fermion field configurations that violate lepton number by two units but missed $u^2d^{*2}e^2$. For the field configuration $Qd^*L^3e^*$, we find two independent operators instead of one as given in that reference. For the configuration $Q^{*2}u^2L^2$, Ref.~\cite{Babu:2001ex} claims 4 operators after considering two different color contractions in the four-quark part, but we only obtain 2 independent operators. In addition, Refs.~\cite{Graesser:2016bpz,He:2020jly} list part of operators contributing to the nuclear and kaon neutrinoless double beta decays, respectively.

\noindent
{$\scriptscriptstyle\blacksquare$ $(\Delta B, \Delta L)=(2,0)$ sector}

The SM gauge symmetry restricts the field configurations in this sector to be $d^2Q^4$, $d^3Q^2u$, and $d^6u^2$. One first uses FIs to fix the scalar structures, and then performs the $SU(2)_L$ contractions. Finally, the color contractions have to be considered carefully due to SIs involving two totally antisymmetric tensors in equation~\eqref{SI:r4}. We find there are 5 operators after considering all algebraic identities which are given at the top of table~\ref{tab:BV1}. In appendix~\ref{app:flavorsymmetry}, we derive flavor relations for each operator to confirm our counting of operators for the $n_f$-generation case. Restricting ourselves to the first generation of fermions, there are 4 operators which contribute potentially to the neutron-antineutron oscillation as considered previously in~\cite{Rao:1983sd,Caswell:1982qs}.

\noindent
{$\scriptscriptstyle\blacksquare$ $(\Delta B, \Delta L)=(1,3)$ sector}

There are only two field configurations $L^3Qu^2$ and $L^2eu^3$ by the SM gauge symmetry, and each of them fits into one scalar structure shown in table~\ref{tab:BV1}. One can see these operators vanish for one generation of fermions, as had been remarked by Weinberg in Ref.~\cite{Weinberg:1980bf}.

\noindent
{$\scriptscriptstyle\blacksquare$ $(\Delta B, \Delta L)=(1,-1)$ sector}

The field configurations in this sector are divided into two cases with either three quarks plus three leptons or five quarks plus one lepton as separated by a thick line in table~\ref{tab:BV2}. For the former case, there are six field configurations which are completely covered by seven generic operators. In Ref.~\cite{Hambye:2017qix} however, the authors listed sixteen operators, many of which we find to be redundant by the use of the first FI in equation~\eqref{FI2}. For the latter case, we have to consider the first SI in equation~\eqref{SI:r3} to exclude redundant operators due to various color contractions. With the compact operator basis given in table~\ref{tab:BV2}, any other operators formed for each field configuration can be transformed into the list by FIs and SIs.

\subsection{Classes $\psi^2\varphi^5D$ and $\psi^2\varphi^3XD$}

The SM gauge symmetry only allows the field configuration $LeH^4H^*D$ in the class $\psi^2\varphi^5D$ which yields a unique operator, while it singles out the field configuration $eLH^3XD$ in the class $\psi^2\varphi^3XD$ with $X$ being either $B$ or $W$. These operators are shown in table~\ref{tab:LV2}. Caution must be practised when constructing operators with $X=W$ as it is necessary to remove redundancy by the identity in equation~\eqref{SIW}.

\subsection{Class $\psi^4\varphi^2D$}

Lorentz invariance apparently requires the four-fermion part to be either a scalar-vector or a tensor-vector combination:
\begin{align}
&{\cal O}_{SV}:~
(\overline{\psi_1}\psi_2)(\overline{\psi_3}\gamma_\mu \psi_4)\varphi^2 D^\mu\;,~
&& {\cal O}_{TV}:~
(\overline{\psi_1}\sigma_{\mu\nu}\psi_2)(\overline{\psi_3}\gamma^\mu \psi_4)\varphi^2 D^\nu\;.
\end{align}
However, the structure ${\cal O}_{TV}$ can be transformed into ${\cal O}_{SV}$ by the following FIs,
\begin{align}\nonumber
i\sigma^{\mu\nu}P_\mp  \otimes \gamma_\nu P_\pm =P_\mp \otimes \gamma^\mu P_\pm +2 \gamma^\mu P_\pm \odot  P_\mp,
\\
i\sigma^{\nu\mu}P_\pm  \otimes \gamma_\nu P_\pm =P_\pm \otimes \gamma^\mu P_\pm +2 P_\pm\odot  \gamma^\mu  P_\pm.
\label{FI5}
\end{align}
It is thus sufficient to focus on the structure ${\cal O}_{SV}$. Now we attach the covariant derivative $D^\mu$ to an appropriate field. With IBP we choose $\psi_1$ to be derivative free, and with EoM we avoid associating $D^\mu$ with the vector current, so that it acts on either $\psi_2$ or the scalar field.  The final operators in this class are listed in table~\ref{tab:LV3} for the $(\Delta B, \Delta L)=(0,2)$ sector and table~\ref{tab:BV2} for the $(\Delta B, \Delta L)=(1,-1)$ sector. In reaching the listed operators for each field configuration, we have employed FIs in equation~\eqref{FI2} and their variants to remove redundant operators arising from the exchange of fermions in the two bilinears. In addition, the SI for the rank-2 totally antisymmetric tensor in equation~\eqref{SI:r2} is also useful to reduce operators due to apparently different $SU(2)_L$ contractions.

We take the field configuration $L^2ee^*H^2D$ in the $(\Delta B,\Delta L)=(0,2)$ sector to illustrate our main points outlined above. Before attaching $D^\mu$ to a specific field, the structure ${\cal O}_{SV}$ appears as,
\begin{align}
\epsilon_{ik}\epsilon_{jl}(\overline{e}L^i)(\overline{e^{\C}}\gamma_\mu L^j) H^kH^l D^\mu.
\end{align}
In this example the $SU(2)_L$ contraction is unique, and the other operator with two $L$s in the same bilinear can be transformed into the above one by the FIs shown in equation \eqref{FI2} and equation \eqref{FI5}. Now we attach $D^\mu$ to a field. By excluding EoM terms and considering IBP relations, we can act $D^\mu$ on $H^k$, $H^l$ or $L^i$. The former two cases lead directly to the two operators shown in table~\ref{tab:LV3}. The latter case can be recast by the second FI in equation \eqref{FI2} into a symmetric form with the two $L$s in the same bilinear. We thus obtain three operators for this field configuration in the general flavor case. In a similar fashion we can work out all other field configurations.

\subsection{Class $\psi^4XD$}

By Lorentz invariance the four-fermion part can be a vector-scalar or vector-tensor combination, so that we have the following four possible structures:
\begin{align}
&{\cal O}_{VS}:~(\overline{\psi_1}\gamma_\mu \psi_2)(\overline{\psi_3^\C}\psi_4) X^{\mu\nu}D_\nu\;,~
&&{\cal O}_{VTa}:~(\overline{\psi_1}\gamma_\mu \psi_2)(\overline{\psi_3^\C}\sigma_{\nu\rho}\psi_4)X^{\nu\rho}D^\mu\;,
\nonumber
\\
&{\cal O}_{VTb}:~(\overline{\psi_1}\gamma_\mu \psi_2)(\overline{\psi_3^\C}\sigma_{\nu\rho}\psi_4)X^{\mu\rho}D^\nu\;,~
&&{\cal O}_{VTc}:~(\overline{\psi_1}\gamma_\mu \psi_2)(\overline{\psi_3^\C}\sigma^{\mu\rho}\psi_4)X_{\nu\rho}D^\nu\;.
\end{align}
The field strength tensor $X$ can also be its dual $\tilde X$ for ${\cal O}_{VS}$, while for ${\cal O}_{VTa,b,c}$ the dual case is reducible to the above four structures by GIs in equation~\eqref{GI}. As a matter of fact, we will show below that all structures ${\cal O}_{VTa,b,c}$ can be reduced to the structure ${\cal O}_{VS}$.

We start with ${\cal O}_{VTc}$, which is easily reduced to the $VS$ structure by FI in equation~\eqref{FI5}. For ${\cal O}_{VTa,b}$, we assume that $\psi_{2,3,4}$ have the same chirality; if this is not the case, we rewrite $(\overline{\psi_1}\gamma_\mu \psi_2)=-(\overline{\psi_2^\C}\gamma_\mu \psi_1^\C)$ so that $\psi_1^\C$ has the same chirality as $\psi_{3,4}$. It is easy to see that there are two operators for ${\cal O}_{VTa}$ that are free of EoM and IBP relations,
\begin{align}
{\cal O}_{VTa1}=(\overline{\psi_1}\gamma_\mu \psi_2)(\overline{D^\mu\psi_3^\C}\sigma_{\nu\rho}\psi_4)X^{\nu\rho}\;,~
{\cal O}_{VTa2}=(\overline{\psi_1}\gamma_\mu \psi_2)(\overline{\psi_3^\C}\sigma_{\nu\rho}D^\mu\psi_4)X^{\nu\rho}\;.
\end{align}
By a generalized FI as in equation~\eqref{FI2}, ${\cal O}_{VTa1}$ can be recast as
\begin{align}
{\cal O}_{VTa1}=-(\overline{\psi_1}\gamma_\mu D^\mu\psi_3)(\overline{\psi_2^\C}\sigma_{\nu\rho}\psi_4)X^{\nu\rho}
-(\overline{\psi_1}\gamma_\mu \sigma_{\nu\rho}\psi_4 )(\overline{\psi_2^\C}D^\mu\psi_3)X^{\nu\rho}\;,
\end{align}
where the first term reduces to EoM operators and the second to $VS$ operators by the GI involving three gamma matrices in equation~\eqref{GI}. A similar manipulation applies to ${\cal O}_{VTa2}$. For ${\cal O}_{VTb}$, when $D^\nu$ acts on $\psi_{3,4}$ it reduces to scalar structures and EoM operators by the GIs in equation~\eqref{GI}. Therefore, with IBP relations, there are also two possible operators,
\begin{align}
&{\cal O}_{VTb1}=(\overline{\psi_1}\gamma_\mu D^\nu\psi_2)(\overline{\psi_3^\C}\sigma_{\nu\rho}\psi_4)X^{\mu\rho}\;,
&&{\cal O}_{VTb2}=(\overline{\psi_1}\gamma_\mu \psi_2)(\overline{\psi_3^\C}\sigma_{\nu\rho}\psi_4)D^\nu X^{\mu\rho}\;.
\end{align}
However, after the use of BI in equation~\eqref{BI} that is followed by the IBP and EoM procedure, ${\cal O}_{VTb2}$ actually reduces to ${\cal O}_{VTa1,a2}$ which are themselves reducible. For ${\cal O}_{VTb1}$, the reduction to the $VS$ structure is similar to ${\cal O}_{VTa1,a2}$. This establishes our claim that all operators in this class are covered by the $VS$ structure ${\cal O}_{VS}$.

We now consider the insertion of a covariant derivative in ${\cal O}_{VS}$. Upon excluding EoM operators and making $\psi_4$ derivative free by IBP, we obtain six possible operators,
\begin{align}
\nonumber
&{\cal O}_{VS1}=(\overline{D_\nu \psi_1}\gamma_\mu \psi_2)(\overline{\psi_3^\C}\psi_4) X^{\mu\nu}\;,~
&&{\cal O}_{VS2}=(\overline{\psi_1}\gamma_\mu D_\nu\psi_2)(\overline{\psi_3^\C}\psi_4)X^{\mu\nu}\;,~
\\
\nonumber
&{\cal O}_{VS3}=(\overline{\psi_1}\gamma_\mu  \psi_2)(\overline{D_\nu\psi_3^\C}\psi_4) X^{\mu\nu}\;,~
&&{\cal O}_{VS4}=(\overline{D_\nu\psi_1}\gamma_\mu  \psi_2)(\overline{\psi_3^\C}\psi_4) \tilde X^{\mu\nu}\;,~
\\
&{\cal O}_{VS5}=(\overline{\psi_1}\gamma_\mu D_\nu\psi_2)(\overline{\psi_3^\C}\psi_4) \tilde X^{\mu\nu}\;,~
&&{\cal O}_{VS6}=(\overline{\psi_1}\gamma_\mu \psi_2)(\overline{D_\nu\psi_3^\C}\psi_4) \tilde X^{\mu\nu}\;.
\end{align}
The operators ${\cal O}_{VS4,5}$ are reducible due to the second GI in equation~\eqref{GI} and by the use of the EoMs of $\psi_{1,2}$. This leaves us with the operators ${\cal O}_{VS1,2,3,6}$ which may have the field configurations $ud^*L^2XD$, $d^3e^*XD$, and $Qd^2L^*XD$. To obtain the minimal set of independent operators, it is necessary to employ the FI in equation~\eqref{FI2} and the SI in equation~\eqref{SI:r2}. Our final results are shown in table~\ref{tab:LV4} for $(\Delta B, \Delta L)=(0,2)$ and table~\ref{tab:BV3} for $(\Delta B, \Delta L)=(1,-1)$ sectors respectively.

\subsection{Class $\psi^2\varphi^4D^2$}

Gauge invariance requires the field configurations in this class to be either $e^2H^4D^2$ or $L^2H^* H^3D^2$. For $e^2H^4D^2$, one readily gets a unique operator as shown in table~\ref{tab:LV4}. For $L^2H^* H^3D^2$, we first consider $SU(2)_L$ contractions which take the unique form $\epsilon_{ik}\epsilon_{jl}(H^\dagger H)L^i L^jH^k H^l D^2$ before inserting covariant derivatives into proper positions. Next, we consider the fermion bilinear which can take either a scalar or a tensor type:
\begin{align}
\epsilon_{ik}\epsilon_{jl}(H^\dagger H) (\overline{L^{\C,i}} L^j)H^k H^l D^\mu D_\mu \;, ~
\epsilon_{ik}\epsilon_{jl}(H^\dagger H) (\overline{L^{\C,i}}\sigma_{\mu\nu} L^j)H^k H^l D^\mu D^\nu \;.
\end{align}
Now we attach covariant derivatives to fields to form specific operators. For the tensor type, the operators with a derivative acting on the lepton field $L$ can be reduced to the scalar ones or EoM operators by the first GI in equation~\eqref{GI} which gives the relation $i\sigma_{\mu\nu}D^\mu L=(D_\nu-\gamma_\nu\slashed{D})L$, and can therefore be discarded. By IBP we can arrange $H^*$ to be derivative free, so that we have two possible non-trivial tensor-type operators,
\begin{align}
\nonumber
&{\cal O}_{T1}=\epsilon_{ik}\epsilon_{jl}(H^\dagger D_\mu H) (\overline{L^{\C,i}}\sigma_{\mu\nu} L^j)D^\nu H^k H^l  \;,~
\\
&{\cal O}_{T2}=\epsilon_{ik}\epsilon_{jl}(H^\dagger H) (\overline{L^{\C,i}}\sigma_{\mu\nu} L^j)D_\mu H^kD_\nu H^l  \;.
\end{align}
However, by successive application of SIs in equation~\eqref{SI:r2} one finds ${\cal O}_{T2}^{pr}=-({\cal O}_{T1}^{pr}+p\leftrightarrow r)$ so that only ${\cal O}_{T1}$ remains as a tensor operator. For the scalar-type operators, after consideration of IBP (again with $H^*$ to be derivative free), EoM and identical fields, we find five independent operators. All of these six operators for this field configuration are also given in table~\ref{tab:LV4}.

\subsection{Class $\psi^2\varphi^2XD^2$}

Gauge invariance requires the field configuration to be $L^2H^2XD^2$ with the field strength tensor $X=B,~W$. Before attaching covariant derivatives to specific fields, we have the following three types of possibly independent structures in terms of fermion bilinears and Lorentz contractions:
\begin{align}
&{\cal O}_{S}:~\epsilon_{ik}(\epsilon X_{\mu\nu})_{jl}(\overline{L^{\C,i}}L^j)H^k H^lD^\mu D^\nu\;,
\nonumber
\\
&{\cal O}_{Ta}:~\epsilon_{ik}(\epsilon X_{\mu\nu})_{jl}(\overline{L^{\C,i}}\sigma^{\mu\nu}L^j)H^k H^lD_\alpha D^\alpha\;,
\nonumber
\\
&{\cal O}_{Tb}:~\epsilon_{ik}(\epsilon X_{\mu\rho})_{jl}(\overline{L^{\C,i}}\sigma_{\nu}^{~\rho}L^j)H^k H^l D^\mu D^\nu\;,
\end{align}
with $(\epsilon B_{\mu\rho})_{jl}=B_{\mu\rho}\epsilon_{jl}$. All other $SU(2)_L$ contractions can be transformed into the above ones by SIs in equation~\eqref{SI:r2}. For the scalar structure ${\cal O}_{S}$, the field strength tensor can also be its dual $\tilde X^{\mu\nu}$, while for the tensor structures ${\cal O}_{Ta,b}$ the cases with a dual field strength tensor can be transformed into ${\cal O}_{Ta,b}$ by GIs involving $\epsilon_{\mu\nu\rho\sigma}$ in equation~\eqref{GI}. In the following we try to keep operators in the scalar structure as many as possible by reducing operators in tensor structures.

We start with ${\cal O}_{Tb}$. With either $D^\mu$ or $D^\nu$ acting on a lepton field it can be reduced into ${\cal O}_{S}$ and/or ${\cal O}_{Ta1}$. This is obvious for $D^\nu$ via the first GI in equation~\eqref{GI} and the EoM for the lepton field. For $D^\mu$ acting on a lepton field, we rewrite $X_{\mu\rho}=-\epsilon_{\mu\rho\alpha\beta}\tilde X^{\alpha\beta}/2$ and then apply the last and first GIs in equation~\eqref{GI} and the EoM for the lepton field. Furthermore, when either $D^\mu$ or $D^\nu$ acts on $X_{\mu\rho}$ it can be shifted away by the EoM or the BI in equation~\eqref{BI} followed by IBP to reduce to the structure ${\cal O}_{Ta}$. In this way, after IBP and EoM we obtain one operator in ${\cal O}_{Tb}$ with each Higgs field being acted upon by one covariant derivative,
\begin{align}
{\cal O}_{Tb1}=\epsilon_{ik}(\epsilon X_{\mu\rho})_{jl}(\overline{L^{\C,i}}\sigma_{\nu}^{~\rho}L^j)
D^\mu H^k D^\nu H^l\;.
\end{align}
Note that when $X_{\mu\rho}=B_{\mu\rho}$, ${\cal O}_{Tb1}$ is antisymmetric under the exchange of the two leptons up to EoM operators and operators covered in ${\cal O}_{S}$ and ${\cal O}_{Ta}$. Next we consider ${\cal O}_{Ta}$. Modulo IBP (with $X_{\mu\nu}$ to be derivative free) and EoM, we have the following six operators in ${\cal O}_{Ta}$:
\begin{align}
&{\cal O}_{Ta1}=\epsilon_{ik}(\epsilon X_{\mu\nu})_{jl}
(\overline{D_\alpha L^{\C,i}}\sigma_{\mu\nu} L^j)D^\alpha H^k H^l\;,~
\nonumber
\\
&{\cal O}_{Ta2}=\epsilon_{ik}(\epsilon X_{\mu\nu})_{jl}
(\overline{D_\alpha L^{\C,i}}\sigma^{\mu\nu} L^j) H^k D^\alpha H^l\;,
\nonumber
\\
&{\cal O}_{Ta3}=\epsilon_{ik}(\epsilon X_{\mu\nu})_{jl}
(\overline{L^{\C,i}}\sigma^{\mu\nu}D_\alpha L^j)D^\alpha H^k H^l\;,~
\nonumber
\\
&{\cal O}_{Ta4}=\epsilon_{ik}(\epsilon X_{\mu\nu} )_{jl}
(\overline{L^{\C,i}}\sigma^{\mu\nu}D_\alpha  L^j) H^kD^\alpha H^l\;,
\nonumber
\\
&
{\cal O}_{Ta5}=\epsilon_{ik}(\epsilon X_{\mu\nu}) _{jl}
(\overline{L^{\C,i}}\sigma^{\mu\nu} L^j)D_\alpha H^kD^\alpha H^l\;,~
\nonumber
\\
&{\cal O}_{Ta6}=\epsilon_{ik}(\epsilon X_{\mu\nu})_{jl}(\overline{D_\alpha L^{\C,i}}\sigma^{\mu\nu}D^\alpha L^j) H^k H^l\;.
\end{align}
Since $\sum_{i}{\cal O}_{Tai}$ contains only EoM terms, ${\cal O}_{Ta5}$ can be excluded. Furthermore, by rewriting $X_{\mu\nu}=-\epsilon_{\mu\nu\rho\sigma}\tilde X^{\rho\sigma}/2$, ${\cal O}_{Ta6}$ can be transformed into ${\cal O}_{S}$ plus EoM operators by the SI in equation~\eqref{SI:r4} and the first GI in equation~\eqref{GI}. For the remaining four operators, one can easily figure out that only ${\cal O}_{Ta1,2}$ are independent for $X_{\mu\nu}=B_{\mu\nu}$, while there are three independent operators for $X_{\mu\nu}=W_{\mu\nu}$ that may be chosen as ${\cal O}_{Ta1,2,3}$ with ${\cal O}_{Ta4}$ being related to ${\cal O}_{Ta1,2,3}$ by the SI identity $\epsilon_{ik}(\epsilon W_{\mu\nu})_{jl}+\epsilon_{jl}(\epsilon W_{\mu\nu})_{ik}=\epsilon_{il}(\epsilon W_{\mu\nu})_{jk}+\epsilon_{jk}(\epsilon W_{\mu\nu})_{il}$ derived from equation~\eqref{SIW}.

At last we consider the scalar structure ${\cal O}_{S}$. Upon applying IBP (with $L^i$ being derivative free) and EoM, one obtains three operators,
\begin{align}
\nonumber
&{\cal O}_{S1}=\epsilon_{ik}(\epsilon X_{\mu\nu})_{jl}
(\overline{L^{\C,i}}D^\mu L^j)D^\nu H^k H^l\;,
\\
&{\cal O}_{S2}=\epsilon_{ik}(\epsilon X_{\mu\nu})_{jl}
(\overline{L^{\C,i}}D^\mu L^j) H^kD^\nu H^l\;,~
\nonumber
\\
&{\cal O}_{S3}=\epsilon_{ik}(\epsilon X_{\mu\nu})_{jl}
(\overline{L^{\C,i}} L^j)D^\mu H^k D^\nu H^l\;,
\end{align}
plus three more with $X_{\mu\nu}$ being replaced by its dual. For $X_{\mu\nu}=B_{\mu\nu}$, ${\cal O}_{S1}$ can be expressed as a sum of ${\cal O}_{S2,3}$ up to EoM operators by IBP manipulation, while for $X_{\mu\nu}=W_{\mu\nu}$ the operator ${\cal O}_{S3}$ can be expressed as a sum of ${\cal O}_{S1,2}$ up to EoM operators by IBP and the above SI manipulation. In summary, we have the independent operators ${\cal O}_{S2,3}$ plus their dual cases, ${\cal O}_{Ta1,2}$, and ${\cal O}_{Tb1}$ for the field configuration $L^2H^2BD^2$, and ${\cal O}_{S1,2}$ plus their dual cases, ${\cal O}_{Ta1,2,3}$, and ${\cal O}_{Tb1}$ for $L^2H^2WD^2$. All of these operators are listed in table~\ref{tab:LV4} with fermion flavors counted.

\subsection{Class $\psi^4\varphi D^2$}

For operators in this class one first exploits the Fierz identities to arrange the two fermion bilinears to be in the scalar or tensor form, so that there are four structures to begin with,
\begin{align}
\nonumber
&{\cal O}_{SS}:~(\overline{\psi_1} \psi_2)(\overline{\psi_3^\C}\psi_4)\varphi D^\mu  D_\mu\;,~
&&{\cal O}_{TS}:~(\overline{\psi_1}\sigma_{\mu\nu} \psi_2)(\overline{\psi_3^\C}\psi_4)\varphi D^\mu  D^\nu\;,
\\
&{\cal O}_{ST}:~(\overline{\psi_1} \psi_2)(\overline{\psi_3^\C}\sigma_{\mu\nu}\psi_4)\varphi D^\mu  D^\nu\;,~
&&{\cal O}_{TT}:~(\overline{\psi_1}\sigma_{\mu\rho} \psi_2)(\overline{\psi_3^\C}\sigma_{\nu}^{~\rho}\psi_4)\varphi D^\mu D^\nu\;,~
\end{align}
where we use $\psi_3^\C$ instead of $\psi_3$ because all operators violate lepton or baryon number or both. Next we attach derivatives to fields to form possible operators. To do that, we note that the tensor-tensor structure can be reduced into others and EoM operators by the following identities
\begin{align}
&\sigma_{\mu\rho}P_\pm\otimes \sigma_{\nu}^{~\rho}P_\mp -\mu\leftrightarrow \nu=0,
\nonumber
\\
&\sigma_{\mu\rho}P_\pm\otimes \sigma_{\nu}^{~\rho}P_\pm +\mu\leftrightarrow \nu=-2g_{\mu\nu}\sigma_{\rho\sigma}P_\pm  \otimes \sigma^{\rho\sigma}P_\pm =8g_{\mu\nu}(P_\pm \otimes P_\pm +2P_\pm\odot P_\pm),
\end{align}
together with IBP (with $\varphi$ to be derivative free), EoMs (for instance, $\overline{\psi_1}i\sigma_{\mu\rho}D^\rho \psi_2=\overline{\psi_1}D_\mu\psi_2-\overline{\psi_1}\gamma_\mu\slashed{D}\psi_2$). Therefore, we only need to consider ${\cal O}_{SS},{\cal O}_{TS},{\cal O}_{ST}$ in the following.

For the scalar-tensor structures ${\cal O}_{TS}$ and ${\cal O}_{ST}$, taking into account IBP (again with $\varphi$ being derivative free) and EoM, we arrive at the following unique forms
\begin{align}
&{\cal O}_{TS}^{XY}=(\overline{\psi_1}\sigma_{\mu\nu}P_X \psi_2)
(\overline{D^\mu \psi_3^\C}D^\nu P_Y\psi_4)\varphi \;,
\nonumber
\\
&{\cal O}_{ST}^{XY}=(\overline{D_\mu \psi_1}D_\nu P_X \psi_2)
(\overline{\psi_3^\C}\sigma^{\mu\nu}P_Y\psi_4)\varphi  \;,~
\end{align}
where we make chiral projectors $P_X$ with $X=\pm$ manifest for convenience. If $\psi_{2,3,4}$ have the same chirality, the operators ${\cal O}_{TS}^{XX},~{\cal O}_{ST}^{XX}$ can be reduced into the scalar-scalar operators ${\cal O}_{SS}$ and EoM operators by the FIs in equation~\eqref{FI2}; for example, for ${\cal O}_{ST}^{\pm\pm}$, we have
\begin{align}
\nonumber
&\quad (\overline{D_\mu \psi_1}D_\nu P_\pm\psi_2)(\overline{\psi_3^\C}\sigma^{\mu\nu}P_\pm\psi_4)\varphi
\\
&=-(\overline{D_\mu \psi_1}\sigma^{\mu\nu}P_\pm\psi_4)(\overline{\psi_3^\C}D_\nu P_\pm\psi_2 )\varphi
-(\overline{D_\mu \psi_1}P_\pm \psi_3)(\overline{D_\nu \psi_2^\C}\sigma^{\mu\nu}P_\pm\psi_4)\varphi\;,
\end{align}
where the tensor currents can be reduced into scalar and EoM terms as we illustrated above. The similar manipulation also applies to ${\cal O}_{TS}^{\pm\pm}$ with $\psi_{2,3,4}$ having the same chirality. Thus for the scalar-tensor structures only ${\cal O}_{TS}^{\pm\mp}$ and ${\cal O}_{TS}^{\pm\mp}$ survive.

Now we come to the scalar-scalar structure ${\cal O}_{SS}$. In this case we can choose to make $\psi_1$ free of derivatives by IBP, then there are generally six ways to distribute two derivatives among the four fields ($\psi_{2,3,4}$ and $\varphi$). But one of the six is still redundant by IBP and EoM. (In the on-shell language this redundancy comes from the momentum relation $(p_2+p_3+p_4+p_\varphi)^2=p_1^2=m_1^2$.) Therefore, we obtain five possible independent forms which can be chosen to be
\begin{align}
\nonumber
&{\cal O}_{SS1}=(\overline{\psi_1}D_\mu \psi_2)(\overline{D^\mu\psi_3^\C}\psi_4)\varphi\;,
&&
\\
\nonumber
&{\cal O}_{SS2}=(\overline{\psi_1}D_\mu \psi_2)(\overline{\psi_3^\C}D^\mu \psi_4)\varphi\;,~
&&{\cal O}_{SS3}=(\overline{\psi_1}D_\mu \psi_2)(\overline{\psi_3^\C}\psi_4)D^\mu\varphi\;,~
\\
&{\cal O}_{SS4}=(\overline{\psi_1} \psi_2)(\overline{D_\mu\psi_3^\C}\psi_4)D^\mu\varphi\;,~
&&{\cal O}_{SS5}=(\overline{\psi_1} \psi_2)(\overline{\psi_3^\C} D_\mu\psi_4)D^\mu\varphi\;.
\end{align}
If there are identical fields among $\psi_{2,3,4}$, the above five forms could be not completely independent and the redundant ones can be easily identified at this stage. At last by combining the field configurations in this class from the HS output with the above analysis of structures, we obtain the final operator basis as shown in table~\ref{tab:LV5} for the $(\Delta B,\Delta L)=(0,2)$ sector and in table~\ref{tab:BV3} for the $(\Delta B,\Delta L)=(1,-1)$ sector. One should be careful with the operators with multiple $SU(2)_L$ fields in table~\ref{tab:LV5}, since the SIs in equation~\eqref{SI:r2} have to be used to reduce any redundant operators coming from apparently different $SU(2)_L$ contractions.

\subsection{Class $\psi^2\varphi^3D^3$}

The only possible field configuration is $eLH^3D^3$ with a vector fermion bilinear by Lorentz invariance. With four Lorentz vector indices at hand, they either contract with the tensor $\epsilon^{\mu\nu\rho\sigma}$ or self-contract in pair. The former yields the unique term $\epsilon^{\mu\nu\rho\sigma}\epsilon_{ij}\epsilon_{kl}(\overline{e^\C}\gamma_\mu L^i)D_\nu H^j D_\rho H^k D_\sigma H^l$ which vanishes by SI. In the latter case, all possible operators are constructed by attaching a pair of contracted derivatives to Lorentz scalar operators formed by $eLH^3D$. The latter has the unique form, $\epsilon_{ij}\epsilon_{kl}(\overline{e^\C}\gamma_\mu L^i)H^j D^\mu H^k H^l$. Modulo the IBP (with the $e$ chosen to be derivative free) and EoM terms, we have the following six possible operators,
\begin{align}
\nonumber
&{\cal O}_1=\epsilon_{ij}\epsilon_{kl}(\overline{e^\C}\gamma_\mu D_\nu L^i)D^\nu H^j D^\mu H^k H^l\;,~
{\cal O}_2=\epsilon_{ij}\epsilon_{kl}(\overline{e^\C}\gamma_\mu D_\nu L^i) H^j D^\nu D^\mu H^k H^l\;,
\\\nonumber
&{\cal O}_3=\epsilon_{ij}\epsilon_{kl}(\overline{e^\C}\gamma_\mu D_\nu L^i) H^j D^\mu H^kD^\nu H^l\;,~
{\cal O}_4=\epsilon_{ij}\epsilon_{kl}(\overline{e^\C}\gamma_\mu L^i)D_\nu H^jD^\nu D^\mu H^k H^l\;,
\\
&{\cal O}_5=\epsilon_{ij}\epsilon_{kl}(\overline{e^\C}\gamma_\mu L^i)D_\nu H^j D^\mu H^kD^\nu H^l\;,~
{\cal O}_6=\epsilon_{ij}\epsilon_{kl}(\overline{e^\C}\gamma_\mu L^i) H^j D_\nu D^\mu H^kD^\nu H^l\;.
\end{align}
Since they sum to EoM terms, we choose to discard ${\cal O}_6$ as redundant. Furthermore, ${\cal O}_{2,4}$ can be reduced as linear combinations of ${\cal O}_{1,3,5}$ by IBP of $D^\mu$ and SIs in equation~\eqref{SI:r2}. Therefore we only have three independent operators in this class which are listed in table~\ref{tab:LV5}.

\subsection{Class $\psi^4D^3$}

The two fermion bilinears in this class can be a vector-scalar or vector-tensor type:
\begin{align}
\nonumber
&{\cal O}_{VS}:~(\overline{\psi_1}\gamma_\mu\psi_2)
(\overline{\psi_3^\C}\psi_4)D^\mu D^2\;,~
\\
&{\cal O}_{VT1}:~(\overline{\psi_1}\gamma^\mu\psi_2)
(\overline{\psi_3^\C}\sigma_{\mu\nu}\psi_4)D^\nu D^2\;,~
\nonumber
\\
&{\cal O}_{VT2}:~(\overline{\psi_1}\gamma_\mu\psi_2)
(\overline{\psi_3^\C}\sigma_{\nu\rho}\psi_4)D^\mu D^\nu D^\rho\;.
\end{align}
One can show that the vector-tensor types may be transformed into the vector-scalar type and EoM operators with the aid of FIs in equation~\eqref{FI2}, GIs in equation~\eqref{GI}, IBP, and EoMs. Modulo IBP (with $\psi_1$ being derivative free) and EoM terms, we have the following possible independent operators of the vector-scalar type,
\begin{align}
{\cal O}_{VS1}=(\overline{\psi_1}\gamma_\mu D_\nu\psi_2)(\overline{D^\nu\psi_3^\C}D^\mu \psi_4)\;,~
{\cal O}_{VS2}=(\overline{\psi_1}\gamma_\mu D_\nu\psi_2)(\overline{D^\mu\psi_3^\C}D^\nu \psi_4)\;.
\end{align}
Taking into account the SM field contents and gauge symmetry, we find that the above two actually condense to a single one for each field configuration. Our results are included in table~\ref{tab:LV5} for the $(\Delta B, \Delta L)=(0,2)$ sector and table~\ref{tab:BV3} for the $(\Delta B, \Delta L)=(1,-1)$ sector, respectively.

\subsection{Class $\psi^2\varphi^2D^4$}

In this class the field configuration can only be $L^2H^2D^4$, in which the lepton bilinear can be a scalar or tensor, with the unique $SU(2)_L$ contractions:
\begin{align}
&{\cal O}_{S}:~\epsilon_{ik}\epsilon_{jl}
(\overline{L^{\C,i}}L^j)H^kH^lD^2 D^2\;,~
&&{\cal O}_{T}:~\epsilon_{ik}\epsilon_{jl}
(\overline{L^{\C,i}}\sigma_{\mu\nu}L^j)H^kH^lD^\mu D^\nu D^2\;.
\end{align}
Upon insertion of derivatives one can easily show that the tensor structure is redundant by IBP, the first GI in equation~\eqref{GI} and EoM for the lepton field. For the scalar type ${\cal O}_{S}$, the two pairs of covariant derivatives can be distributed uniformly among the four fields with proper Lorentz contractions. After excluding IBP and EoM operators, we obtain the three operators shown in table~\ref{tab:LV5}.

At the end of this section, we compare our results to those available in the literature and discuss briefly their possible phenomenological consequences. All dim-9 operators in the SMEFT satisfy $|\Delta B-\Delta L|=2$ with $|\Delta B|\le 2$ and $|\Delta L|\le 3$. We discuss them sector by sector according to the values of $(\Delta B,\Delta L)$. In the sector $(\Delta B,\Delta L)=(0,2)$, Refs.~\cite{Babu:2001ex,deGouvea:2007qla,Graesser:2016bpz} listed the subsets of dim-9 operators without involving a covariant derivative $D_\mu$ or gauge field strength tensor $X_{\mu\nu}$, and considered their implications for the generation of neutrino mass and nuclear $0\nu\beta\beta$ decays. Here we provide a basis of complete and independent operators by removing redundant operators in the literature and including additional operators containing a $D_\mu$ or $X_{\mu\nu}$. The latter operators may induce neutrino mass in a novel manner and deserve further investigation. For instance, the operators ${\cal O}_{LLH^2G^21}$ in the class $\psi^2\varphi^2X^2$ and ${\cal O}_{dLQLHG1}$, ${\cal O}_{dQLLHG}$, ${\cal O}_{QuLLHG1,2}$ in the class $\psi^4\varphi X$ may induce neutrino mass via the QCD condensations, $\langle\Omega|G_{\mu\nu}G^{\mu\nu}|\Omega\rangle$ and $\langle\Omega|\bar q G^{\mu\nu}\sigma_{\mu\nu}q|\Omega\rangle$ with $q$ being the up or down quark and $\Omega$ the QCD vacuum. As another example, Ref.~\cite{Gustafsson:2020bou} recently studied the generation of neutrino mass through the operator ${\cal O}_{eeH^4D^2}$ in the class $D^2\psi^2\varphi^4$.

There is a controversy in the literature concerning the matching of effective interactions of dim-9 operators in the SMEFT to those of dim-9 operators in the low-energy effective field theory (LEFT) that contribute to nuclear $0\nu\beta\beta$ decays. There are 24 such operators in the LEFT that respect the $SU(3)_C\times U(1)_{\rm EM}$ gauge symmetry~\cite{Pas:2000vn,Prezeau:2003xn,Graesser:2016bpz,Liao:2019gex}. While Ref.~\cite{Graesser:2016bpz} claimed that only 11 of them can be generated from dim-9 operators in the class $\psi^6$ of the SMEFT and the remaining 13 operators have to be first generated at dimension 11 and 13, Ref.~\cite{Cirigliano:2018yza} found that 8 more operators can also be generated at dimension 9. We confirmed the latter finding. In our basis, the operators that contribute by a tree-level matching to the 19 dim-9 operators in the LEFT are,
\begin{eqnarray}
\nonumber
(a): &&{\cal O}_{ddueue}^\dagger\;,
{\cal O}_{dQdueL1,2}^\dagger\;,
{\cal O}_{QudueL1,2}^\dagger\;,
{\cal O}_{dQdQeL1,2}^\dagger\;,
{\cal O}_{dQQuLL1,2}^\dagger\;,
{\cal O}_{QuQuLL1,2}^\dagger\;;
\\
\nonumber
(b): &&{\cal O}_{LLH^4W1}^\dagger\;,
{\cal O}_{deueH^2D}^\dagger\;,
{\cal O}_{dLuLH^2D2}^\dagger\;,
{\cal O}_{duLLH^2D}^\dagger\;,
{\cal O}_{dQLeH^2D2}^\dagger\;,
{\cal O}_{dLQeH^2D1}^\dagger\;,
\\
&&{\cal O}_{deQLH^2D}^\dagger\;,
{\cal O}_{QueLH^2D2}^\dagger\;,
{\cal O}_{QeuLH^2D2}^\dagger\;,
{\cal O}_{QLQLH^2D2,5}^\dagger\;,
{\cal O}_{QQLLH^2D2}^\dagger\;,
\nonumber
\\
&&{\cal O}_{eeH^4D^2}^\dagger\;,
{\cal O}_{LLH^4D^23,4}^\dagger\;.
\end{eqnarray}
The 11 operators in the group $(a)$ are those already pointed out in~\cite{Graesser:2016bpz} which generate the 11 operators in the LEFT, while the operators in the group $(b)$ all involve a $D_\mu$ or $X_{\mu\nu}$ and yield the 8 additional operators in the LEFT as found in~\cite{Cirigliano:2018yza}. We note in passing that the operators with a $D_\mu$ or $X_{\mu\nu}$ can also lead to a neutrinoless final state in nuclei with 3 electrons and 1 positron, i.e., $0\nu 3e^-1e^+$. Besides nuclear $0\nu\beta\beta$ decays, the $\mu^-\to e^+$ conversion in nuclei gets more attention in recent years~\cite{Geib:2016atx,Geib:2016daa,Berryman:2016slh} and will be explored in the future Mu2e experiment~\cite{Bernstein:2019fyh}. Our basis of complete and independent operators complements the work in~\cite{Berryman:2016slh} and provides a starting point for the investigation of all these processes as well as the lepton-number-violating decays of the mesons, baryons, and the $\tau$ lepton.

In the sector with $(\Delta B, \Delta L)=(2,0)$ we confirm that there are 4 operators contributing to the $n-\bar n$ oscillation~\cite{Caswell:1982qs}. We note that these operators also contribute to the dinucleon transitions in nuclei, $nn\to P_1^+ P_2^-,~P_1^0P_2^0,~\ell^+\ell^-,~\nu\bar\nu,~\gamma\gamma$, $pp\to P_1^+P_2^+$, and $np\to P^+_1P^0_2,~\ell^+\nu$, where $P_{1,2}$ are the lightest pseudoscalar mesons $(\pi, K)$ and $\ell=e,~\mu$. Such processes have clear experimental signatures, and some of them have been constrained by the earlier Frejus and  IMB-3 experiments~\cite{Berger:1991fa,McGrew:1999nd} and the recent Super-K result~\cite{Sussman:2018ylo}. These operators can also induce the oscillation of neutral baryons and antibaryons that involve heavier $s,~c,~b$ quarks. The operators with $(\Delta B,\Delta L)=(1,3)$ first appear at dimension 9 in the SMEFT. As was found long ago~\cite{Weinberg:1980bf}, there are only two operators, but they disappear with one generation of up-type quarks and thus cannot contribute to novel proton decays such as $p\to\ell^+\bar\nu\bar\nu$. Nevertheless, they contribute to exotic heavy baryon decays like $\Lambda_c^+/\Sigma_c^+\to\ell^+\bar\nu_1\bar\nu_2$ and $\Sigma_c^{++}\to\ell^+_1\ell^+_2\bar\nu$. The nucleon decays with $(\Delta B, \Delta L)=(1,-1)$ have been studied previously in Refs.~\cite{Hambye:2017qix,Heeck:2019kgr} from operators in the $\psi^4\varphi^3$ and $\psi^6$ classes. We reduce their operators to a minimal set and in addition include the operators with a $D_\mu$ or $X_{\mu\nu}$. For instance, we observe that the novel decay mode $n\to\nu\gamma$, also constrained by the Super-K experiment~\cite{Takhistov:2015fao}, can be mediated locally at leading order by the operators ${\cal O}_{LdudHB,W}$, ${\cal O}_{LuddHB,W}$, and ${\cal O}_{LdQQHB,W}$ that were missed in previous analyses. Furthermore, the five-body decays $p\to 2\nu\ell^+_1\ell^+_2\ell^-_3$ and $n\to \nu\ell^+_1\ell^+_2 \ell^-_3\ell^-_4$ also first appear from those operators containing a $D_\mu$ or $X_{\mu\nu}$. It would be interesting to investigate these novel processes systematically in the EFT framework to search for imprint of new physics.

\section{Conclusion}
\label{sec4}

We have investigated systematically dim-9 operators in the standard model effective field theory thus pushing the frontier in this direction one step further. Due to the complexity of the issue we have employed two approaches for crosschecks. We applied the Hilbert series to generate the field configurations allowed by Lorentz and gauge symmetries and to count the total number of independent operators that can be formed for each configuration. We then constructed all possible operators explicitly for each field configuration, and made use of all available kinematic and algebraic relations to remove redundant operators. The relations include integration by parts, equations of motion, Schouten identities, Dirac gamma matrix relations and Fierz identities, and Bianchi identities. We analyzed flavor symmetry relations among operators that are applied to remove redundant operators and count complete and independent operators independently of the Hilbert series. The two approaches yield a consistent answer. The basis of dim-9 operators is shown in tables~\ref{tab:LV1}-\ref{tab:BV3}, and can be summarized as follows. All dim-9 operators violate lepton and baryon numbers in various combinations and are thus non-Hermitian. Without counting flavor or generation there are $384|^{\Delta L=\pm 2}_{\Delta B=0}+10|^{\Delta L=0}_{\Delta B=\pm 2}+4|^{\Delta L=\pm3}_{\Delta B=\pm1}+236|^{\Delta L=\mp 1}_{\Delta B=\pm1}$ dim-9 operators; if three generations of fermions in SM are taken into account, there are then $44874|^{\Delta L=\pm 2}_{\Delta B=0}+2862|^{\Delta L=0}_{\Delta B=\pm 2}+486|^{\Delta L=\pm3}_{\Delta B=\pm1}+42234|^{\Delta L=\pm 1}_{\Delta B=\mp1}$ operators. Our result provides a solid starting point for phenomenological analysis. Compared to lower dimensional operators, dim-9 operators can also violate lepton and baryon numbers in the new combinations $\Delta L=0,~\Delta B=\pm 2$ and  $\Delta L=\pm 3,~\Delta B=\pm 1$. The first combination results in the $n-\bar n$ oscillation as studied in Refs.~\cite{Rao:1983sd,Caswell:1982qs,Rinaldi:2018osy}, while the second one can cause unusual baryon decays. We hope to come back to this phenomenological aspect of the issue in the future.

\section*{Acknowledgement}

This work was supported in part by the Grants No.~NSFC-11975130, No.~NSFC-12035008, by The National Key Research and Development Program of China under Grant No. 2017YFA0402200, by the CAS Center for Excellence in Particle Physics (CCEPP), and by the MOST (Grants No. MOST 109-2811-M-002-535, No. MOST 106-2112-M-002-003-MY3). We thank Jiang-Hao Yu for sending us their manuscript~\cite{Li:2020xlh} on the same subject prior to submission and for electronic communications via WeChat. XDM would like to thank J. de Vries for bringing their paper~\cite{Cirigliano:2018yza} into our attention on the issue of effective operators contributing to $0\nu\beta\beta$ processes.

\appendix
\section{Flavor relations}
\label{app:flavorsymmetry}

In this appendix we give the flavor relations for the operators with symmetries in the basis. These relations further confirm our counting of independent operators when fermion flavors are taken into account as in the HS approach. We denote fermion flavors by the Latin letters $(p,r,s,t,v,w)$ of an operator in the same order that fermion fields appear in the operator. The results are summarized first in sectors and then class by class.

\noindent
$\scriptscriptstyle\blacksquare$ $(\Delta B, \Delta L)=(0,2)$ sector

\noindent
Class $\psi^2\varphi^6$:
\begin{align}
{\cal O}_{LLH^6}^{pr}-p\leftrightarrow r=0\;.
\end{align}
Class $\psi^2\varphi^4X$:
\begin{align}
&{\cal O}_{LLH^4B}^{pr}+p\leftrightarrow r=0\;,~
&&{\cal O}_{LLH^4W2}^{pr}+p\leftrightarrow r=0\;.
\end{align}
Class $\psi^2\varphi^2X^2$ ($x=1,2$ and $y=1,\dots,5$):
\begin{align}\nonumber
&{\cal O}_{LLH^2B^2x}^{pr}-p\leftrightarrow r=0\;,~
&&{\cal O}_{LLH^2W^2y}^{pr}-p\leftrightarrow r=0\;,~
\\
&{\cal O}_{LLH^2W^26}^{pr}+p\leftrightarrow r=0\;,~
&&{\cal O}_{LLH^2G^2x}^{pr}-p\leftrightarrow r=0\;.
\end{align}
Class $\psi^4\varphi X$:
\begin{align}\nonumber
&{\cal O}_{eLLLHW2}^{prst}+s\rightarrow t=0\;,~
&&{\cal O}_{dQLLHW2}^{prst}+s\rightarrow t=0\;,~
\\
&{\cal O}_{QuLLHW2}^{prst}-s\rightarrow t=0\;,~
&&{\cal O}_{QuLLHW4}^{prst}+s\rightarrow t=0\;.
\end{align}
Class $\psi^4\varphi^3$:
\begin{align}
&{\cal O}_{LeLLH^3}^{prst}-s\leftrightarrow t=0\;,~
&&{\cal O}_{QuLLH^32}^{prst}-s\leftrightarrow t=0\;,~
&&{\cal O}_{QdLLH^3}^{prst}-s\leftrightarrow t=0\;.
\end{align}
Class $\psi^6$ ($x=1,2$ and $y=1,2,4$):
\begin{align}\nonumber
&{\cal O}_{eLeLLL}^{prstvw}-{\cal O}_{eLeLLL}^{stprwv}=0\;,
&&\big({\cal O}_{eLeLLL}^{prstvw}+t \leftrightarrow v\big) - t \leftrightarrow w=0\;,
\\\nonumber
&\big({\cal O}_{eLeLLL}^{prstvw}+r \leftrightarrow w\big) - r \leftrightarrow v=0\;,
&&\big({\cal O}_{LudLLL}^{prstvw}+t \leftrightarrow v\big) -t \leftrightarrow w=0\;,
\\\nonumber
&\big({\cal O}_{dQeLLL}^{prstvw}+t \leftrightarrow v\big) - t \leftrightarrow w=0\;,
&&\big({\cal O}_{QueLLL}^{prstvw}+t \leftrightarrow v\big) -t \leftrightarrow w=0\;,
\\\nonumber
&{\cal O}_{ddueue}^{prstvw}- {\cal O}_{ddueue}^{rpvwst}=0\;,
&&{\cal O}_{duddLL1}^{prstvw}+v \leftrightarrow w=0\;,~
{\cal O}_{duddLL2}^{prstvw}-v \leftrightarrow w=0\;,
\\\nonumber
&{\cal O}_{duuuLL1}^{prstvw}+v \leftrightarrow w=0\;,
&&{\cal O}_{duuuLL2}^{prstvw}-v \leftrightarrow w=0\;,
\\
&{\cal O}_{dQdQLLy}^{prstvw}-{\cal O}_{dQdQLLy}^{stprwv}=0\;,
&&{\cal O}_{QuQuLLx}^{prstvw}-{\cal O}_{QuQuLLx}^{stprwv}=0\;.
\end{align}
Class $\psi^4\varphi^2D$:
\begin{align}\nonumber
&{\cal O}_{eeLLH^2D}^{prst}+s\leftrightarrow t=0\;,
&&\big({\cal O}_{LLLLH^2D2}^{prst}+r\leftrightarrow s\big)-r\leftrightarrow t=0\;,
\\\nonumber
&{\cal O}_{LLLLH^2D3}^{prst}+s\leftrightarrow t=0\;,
&&{\cal O}_{ddLLH^2D}^{prst}+s\leftrightarrow t=0\;,
\\
&{\cal O}_{uuLLH^2D}^{prst}+s\leftrightarrow t=0\;,
&&{\cal O}_{QQLLH^2D}^{prst}+s\leftrightarrow t=0\;.
\end{align}
Class $\psi^4XD$:
\begin{align}
&{\cal O}_{duLLBD}^{prst}-s\leftrightarrow t=0\;,
&&{\cal O}_{duLLWD}^{prst}+s\leftrightarrow  t=0\;,
&&{\cal O}_{duLLGD}^{prst}-s\leftrightarrow  t=0\;.
\end{align}
Class $\psi^2\varphi^4D^2$:
\begin{align}
&{\cal O}_{eeH^4D^2}^{pr}-p\leftrightarrow r=0\;,~
&&{\cal O}_{LLH^4D^23}^{pr}-p\leftrightarrow r=0\;.
\end{align}
Class $\psi^2\varphi^2XD^2$ ($x=2,4$):
\begin{align}
{\cal O}_{LLH^2D^2Bx}^{pr}+p\leftrightarrow r=0\;,~
{\cal O}_{LLH^2D^2B7}^{pr}+p\leftrightarrow r=\fbox{EoM}+{\cal O}_{LLH^2D^2B1-6} {\rm~terms}\;,
\end{align}
where \fbox{EoM} stands for operators produced via the use of EoM that are already covered in the basis of dim-9 or lower dimensional operators.

\noindent
Class $\psi^4\varphi D^2$:
\begin{align}
{\cal O}_{eLLLH^24}^{prst}+{\cal O}_{eLLLH^24}^{psrt}+{\cal O}_{eLLLH^24}^{ptrs}=\fbox{EoM}+{\cal O}_{eLLLH^21-3} {\rm~terms}\;.
\end{align}
Class $\psi^2\varphi^2 D^4$ ($x=1,2,3$):
\begin{align}
&{\cal O}_{LLH^2D^4x}^{pr}-p\leftrightarrow r=0\;.
\end{align}
\noindent
$\scriptscriptstyle\blacksquare$ $(\Delta B, \Delta L)=(2,0)$ sector
\begin{align}\nonumber
&{\cal O}_{ddQQQQ1}^{prstvw}+{\cal O}_{ddQQQQ1}^{prvwst}=0\;,~
{\cal O}_{ddQQQQ1}^{prstvw}+{\cal O}_{ddQQQQ1}^{rptswv}=0\;,
\\\nonumber
&{\cal O}_{ddQQQQ2}^{prstvw}+{\cal O}_{ddQQQQ2}^{prwvts}=0\;,~
{\cal O}_{ddQQQQ2}^{prstvw}+{\cal O}_{ddQQQQ2}^{rptswv}=0\;,
\\\nonumber
&\big({\cal O}_{ddQQQQ1}^{prstvw}+t\leftrightarrow w\big)-s\leftrightarrow v=0\;,
\\\nonumber
&\big({\cal O}_{ddQQQQ2}^{prstvw}+s\leftrightarrow w\big)-t\leftrightarrow v=
2{\cal O}_{ddQQQQ1}^{prwtsv}\;,
\\\nonumber
&\big({\cal O}_{ddQQQQ2}^{prstvw}+{\cal O}_{ddQQQQ2}^{prsvwt}-{\cal O}_{ddQQQQ2}^{prswvt}\big)-p\leftrightarrow r=
\big({\cal O}_{ddQQQQ1}^{prstvw}+t\leftrightarrow v\big)-p\leftrightarrow r\;,
\\\nonumber
& \big[ \big({\cal O}_{udddQQ}^{prstvw}+ r\leftrightarrow s\big)- s\leftrightarrow t \big] -v\leftrightarrow w=0\;,
\\\nonumber
&{\cal O}_{ududdd1}^{prstvw}+{\cal O}_{ududdd1}^{stprvw}=0\;,~
{\cal O}_{ududdd2}^{prstvw}+{\cal O}_{ududdd2}^{stprwv}=0
\\\nonumber
&\big({\cal O}_{ududdd1}^{prstvw}+{\cal O}_{ududdd1}^{ptswvr}+{\cal O}_{ududdd1}^{pwsrvt}\big)-r\leftrightarrow t=0\;,
\\\nonumber
&\big({\cal O}_{ududdd1}^{prstvw}+{\cal O}_{ududdd1}^{pvstrw}+{\cal O}_{ududdd1}^{prsvtw}\big)-v\leftrightarrow w=0\;,
\\
&\big({\cal O}_{ududdd2}^{prstvw}+t\leftrightarrow w\big)-t\leftrightarrow v= \big({\cal O}_{ududdd1}^{prsvwt}+v\leftrightarrow w\big)-t\leftrightarrow v\;,
\end{align}
\noindent
$\scriptscriptstyle\blacksquare$ $(\Delta B, \Delta L)=(1,3)$ sector
\begin{align}\nonumber
&\big( \calO_{LLLQuu}^{prstvw}+r\leftrightarrow s\big)-p\leftrightarrow s=0\;,
&&\calO_{LLLQuu}^{prstvw}+v\leftrightarrow w=0\;,
\\\nonumber
&\calO_{LLeuuu}^{prstvw}+p\leftrightarrow r=0\;, ~
&&\calO_{LLeuuu}^{prstvw}+v\leftrightarrow w=0\;, ~
\\
&\calO_{LLeuuu}^{prstvw}+\calO_{LLeuuu}^{prsvwt}+\calO_{LLeuuu}^{prswtv}=0\;.
&&
\end{align}
\noindent
$\scriptscriptstyle\blacksquare$ $(\Delta B, \Delta L)=(1,-1)$ sector

\noindent
Class $\psi^4\varphi X$:
\begin{align}\nonumber
&{\cal O}_{LdddHB}^{prst}-s\leftrightarrow t=0\;, ~
&&{\cal O}_{LdddHW}^{prst}-s\leftrightarrow t=0\;,
\\\nonumber
&{\cal O}_{LuddHB}^{prst}-s\leftrightarrow t=0\;, ~
&&{\cal O}_{LuddHW}^{prst}-s\leftrightarrow t=0\;,
\\\nonumber
&{\cal O}_{eQddHB1}^{prst}+s\leftrightarrow t=0\;, ~
&&{\cal O}_{eQddHB2}^{prst}-s\leftrightarrow t=0\;,
\\\nonumber
&{\cal O}_{eQddHW1}^{prst}+s\leftrightarrow t=0\;, ~
&&{\cal O}_{eQddHW2}^{prst}-s\leftrightarrow t=0\;,
\\
&{\cal O}_{LdQQHW2}^{prst}+s\leftrightarrow t=0\;, ~
&&{\cal O}_{LdQQHW4}^{prst}-s\leftrightarrow t=0\;.
\end{align}
Class $\psi^4\varphi^3$:
\begin{align}\nonumber
&{\cal O}_{LdddH^3}^{prst}+s\leftrightarrow t=0\;,
&&{\cal O}_{LdddH^3}^{prst}+{\cal O}_{LdddH^3}^{pstr}+{\cal O}_{LdddH^3}^{ptrs}=0\;,
\\\nonumber
&{\cal O}_{eQddH^3}^{prst}+s\leftrightarrow t=0\;,
&&{\cal O}_{LdQQH^32}^{prst}+s\leftrightarrow t=0\;,~
{\cal O}_{LuQQH^3}^{prst}+s\leftrightarrow t=0\;,
\\
&{\cal O}_{eQQQH^3}^{prst}+s\leftrightarrow t=0\;,~
&&{\cal O}_{eQQQH^3}^{prst}+{\cal O}_{eQQQH^3}^{pstr}+{\cal O}_{eQQQH^3}^{ptrs}=0\;.
\end{align}
Class $\psi^6$:
\begin{align}\nonumber
&{\cal O}_{eeeddd}^{prstvw} - p \leftrightarrow r=0\;,
&&{\cal O}_{eeeddd}^{prstvw}+v\leftrightarrow w=0\;,
\\\nonumber
&{\cal O}_{eeeddd}^{prstvw}+{\cal O}_{eeeddd}^{prsvwt}+{\cal O}_{eeeddd}^{prswtv}=0\;,
&&{\cal O}_{eLLddd}^{prstvw}+v\leftrightarrow w=0\;,
\\\nonumber
&{\cal O}_{eLLddd}^{prstvw}+{\cal O}_{eLLddd}^{prsvwt}+{\cal O}_{eLLddd}^{prswtv}=0\;,
&&{\cal O}_{LLeudd}^{prstvw}- p \leftrightarrow r=0\;,
\\\nonumber
&{\cal O}_{LLeudd}^{prstvw}- v \leftrightarrow w=0\;,
&&{\cal O}_{eddddd}^{prstvw}+v \leftrightarrow w=0\;,
\\\nonumber
&{\cal O}_{eddddd}^{prstvw}-{\cal O}_{eddddd}^{prtsvw}+{\cal O}_{eddddd}^{prvwst}-{\cal O}_{eddddd}^{prwvst}=0\;,
&&{\cal O}_{eddddd}^{prstvw}+{\cal O}_{eddddd}^{prsvwt}+{\cal O}_{eddddd}^{prswtv}=0\;,
\\
&{\cal O}_{LQdddu}^{prstvw}- t \leftrightarrow v=0\;,
&&{\cal O}_{eQuQdd1}^{prstvw}+v \leftrightarrow w=0\;.
\end{align}
Class $\psi^4\varphi^2 D$:
\begin{align}\nonumber
&{\cal O}_{edddH^2D1}^{prst}+{\cal O}_{edddH^2D1}^{pstr}+{\cal O}_{edddH^2D1}^{ptrs}=0\;,
&&{\cal O}_{edddH^2D1}^{prst}+s\leftrightarrow t=0\;,
\\\nonumber
& {\cal O}_{edddH^2D2}^{prst}-s\leftrightarrow t=0\;,
&&{\cal O}_{LQddH^2D1}^{prst}-s\leftrightarrow t=0\;,
\\\nonumber
&{\cal O}_{LQddH^2D2}^{prst}-s\leftrightarrow t=0\;,
&&{\cal O}_{edQQH^2D}^{prst}-s\leftrightarrow t=0\;,
\\
&\big({\cal O}_{LQQQH^2D2}^{prst}-r\leftrightarrow s\big)+r\leftrightarrow t=0\;,
&&{\cal O}_{LQQQH^2D3}^{prst}-s\leftrightarrow t=0\;.
\end{align}
Class $\psi^4XD$:
\begin{align}\nonumber
&{\cal O}_{edddBD2}^{prst}-r\leftrightarrow s=\fbox{EoM}+{\cal O}_{edddBD1}{\rm~terms}\;,
\\\nonumber
&{\cal O}_{edddBD2}^{prst}-s\leftrightarrow t=\fbox{EoM}+{\cal O}_{edddBD1}{\rm~terms}\;,
\\\nonumber
&{\cal O}_{edddGD3}^{prst}-r\leftrightarrow s=\fbox{EoM}+{\cal O}_{edddGD1-2}{\rm~terms}\;,
\\\nonumber
&{\cal O}_{edddGD3}^{prst}+{\cal O}_{edddGD3}^{pstr}+{\cal O}_{edddGD3}^{ptrs}=\fbox{EoM}+{\cal O}_{edddGD1-2}{\rm~terms}\;,
\\
&{\cal O}_{LQddBD}^{prst}-s\leftrightarrow t=0\;,~
{\cal O}_{LQddWD}^{prst}-s\leftrightarrow t=0\;.
\end{align}
Class $\psi^4\varphi D^2$:
\begin{align}\nonumber
&\big( {\cal O}_{LdddHD^22}^{prst}+r\leftrightarrow t\big)-s\leftrightarrow t=\fbox{EoM}+{\cal O}_{LdddHD^21}{\rm~terms}\;,
\\
&{\cal O}_{eQddHD^22}^{prst}+s\leftrightarrow t=0\;,~
{\cal O}_{eQddHD^24}^{prst}-s\leftrightarrow t=0\;,~
{\cal O}_{eQddHD^25}^{prst}-s\leftrightarrow t=0\;.
\end{align}
Class $\psi^4D^3$:
\begin{align}
{\cal O}_{edddD^3}^{prst}+{\cal O}_{edddD^3}^{prts}+{\cal O}_{edddD^3}^{ptrs}=\fbox{EoM}\;.
\end{align}

\newpage
\begin{table}[H]
\centering
\resizebox{\linewidth}{!}{
\renewcommand{\arraystretch}{0.9}
\begin{tabular}{|l |c| l |c|}
\hline
Classes & Configurations & Operators &\# with $ n_f$
\\\hline\hline
\multicolumn{4}{|c|}{{\color{magenta}$(\Delta B, \Delta L)=(0,2)$}}
\\
\hline
\multirow{1}{*}{$\psi^2\varphi^6$ }
& $L^2H^4H^{*2}$
& ${\cal O}_{LLH^6}=\epsilon_{ik}\epsilon_{jl}(\overline{L^{\C,i }}L^j)H^kH^l (H^\dagger H)^2$
& $~~~~~{1\over 2}n_f (n_f+1)~~~~~$
\\
\hline
\multirow{3}{*}{$\psi^2\varphi^4X$}
& $L^2H^3H^{*}B$ 
& ${\cal O}_{LLH^4B}=g_1\epsilon_{ik}\epsilon_{jl}(\overline{L^{\C,i }}\sigma_{\mu\nu}L^j)H^kH^lB^{\mu\nu} (H^\dagger H)$
& ${1\over 2}n_f (n_f-1)$
\\\cline{2-4}
& \multirow{2}{*}{$L^2H^3H^{*}W$ }
& ${\cal O}_{LLH^4W1}=g_2\epsilon_{ik}(\epsilon W^{\mu\nu})_{jl}(\overline{L^{\C,i }}\sigma_{\mu\nu}L^j)H^kH^l(H^\dagger H)$
& $n_f^2$
\\\cline{3-4}
&
& ${\cal O}_{LLH^4W2}=g_2\epsilon_{ik}\epsilon_{jl}(\overline{L^{\C,i }}\sigma_{\mu\nu}L^j)H^kH^l(H^\dagger W^{\mu\nu}H)$
& ${1\over 2}n_f (n_f-1)$
\\\hline
\multirow{13}{*}{$\psi^2\varphi^2X^2$} 
& \multirow{2}{*}{$L^2H^2B^2$ }
& ${\cal O}_{LLH^2B^21}=\alpha_1 \epsilon_{ik}\epsilon_{jl}(\overline{L^{\C,i }}L^j)H^kH^lB_{\mu\nu}B^{\mu\nu}$
& ${1\over 2}n_f (n_f+1)$
\\\cline{3-4}
&
& ${\cal O}_{LLH^2B^22}=\alpha_1 \epsilon_{ik}\epsilon_{jl}(\overline{L^{\C,i }}L^j)H^kH^lB_{\mu\nu}\tilde B^{\mu\nu}$
& ${1\over 2}n_f (n_f+1)$
\\
\hline
\end{tabular}
}
\caption{Dim-9 operators in classes $\psi^2\varphi^6,~\psi^2\varphi^4X,~\psi^2\varphi^2X^2,~\psi^4\varphi X$ for sector $(\Delta B,\Delta L)=(0,2)$. There are 45 operators without counting fermion generations and 39 remain for one generation case. Hermitian conjugated operators are not included or counted.}
\label{tab:LV1}
\end{table}

\begin{table}[H]
\centering
\resizebox{\linewidth}{!}{
\renewcommand{\arraystretch}{1.2}
\begin{tabular}{|l |c| l |c|}
\hline
Classes & Configurations & Operators &\# with $ n_f$
\\\hline\hline
\multicolumn{4}{|c|}{{\color{magenta}$(\Delta B, \Delta L)=(0,2)$}}
\\\hline
\multirow{11}{*}{$\psi^4\varphi^3$}
& \multirow{1}{*}{$L^3e^* H^2H^*$}
& ${\cal O}_{eLLLH^3}=\epsilon_{ik}\epsilon_{jl}\epsilon_{mn}(\overline{e}L^i)(\overline{L^{\C,j}}L^m)H^kH^l \tilde H^n$
& $n_f^4$
\\\cline{2-4}
& $L^2L^*e H^3$
& ${\cal O}_{LeLLH^3}=\epsilon_{ik}\epsilon_{jl}(\overline{L}eH)(\overline{L^{\C,i }}L^j)H^kH^l$
& ${1\over 2}n_f^3(n_f+1)$
\\\clineB{2-4}{3}
& \multirow{1}{*}{$ud^*LeH^2H^*$}
& ${\cal O}_{dLueH^3}=\epsilon_{ij}(\overline{d}L^i)(\overline{u^\C}e)H^j(H^\dagger H)$
& $n_f^4$
\\\cline{2-4}
& \multirow{3}{*}{$Qd^*L^2H^2H^*$}
& ${\cal O}_{dLQLH^31}=\epsilon_{ik}\epsilon_{jl}\epsilon_{mn}(\overline{d}L^m)(\overline{Q^{\C,i}}L^j)H^kH^l \tilde H^n$
& $n_f^4$
\\\cline{3-4}
&
& ${\cal O}_{dLQLH^32}=\epsilon_{ik}\epsilon_{jl}\epsilon_{mn}(\overline{d}L^i)(\overline{Q^{\C,m}}L^j)H^kH^l \tilde H^n$
& $n_f^4$
\\\cline{3-4}
&
& ${\cal O}_{dLQLH^33}=\epsilon_{ik}\epsilon_{jl}\epsilon_{mn}(\overline{d}L^i)(\overline{Q^{\C,j}}L^m)H^kH^l \tilde H^n$
& $n_f^4$
\\
\hline
\end{tabular}
}
\caption{Dim-9 operators in classes $\psi^4\varphi^3,~\psi^6,~D\psi^2\varphi^5, ~D\psi^2\varphi^3X$ for sector $(\Delta B,\Delta L)=(0,2)$. There are 43 operators without counting fermion generations and 41 remain for one generation case. Hermitian conjugated operators are not included or counted.}
\label{tab:LV2}
\end{table}

\begin{table}[H]
\centering
\resizebox{\linewidth}{!}{
\renewcommand{\arraystretch}{1.2}
\begin{tabular}{|l |c| l |c|}
\hline
Classes & Configurations & Operators &\# with $ n_f$
\\\hline\hline
\multicolumn{4}{|c|}{{\color{magenta}$(\Delta B, \Delta L)=(0,2)$}}
\\
\hline
\multirow{35}{*}{$D\psi^4\varphi^2$}
&\multirow{3}{*}{$L^2ee^*H^2D$}
& ${\cal O}_{eLeLH^2D1}=\epsilon_{ik}\epsilon_{jl}(\overline{e}L^i)(\overline{e^{\C}}\gamma_\mu L^j)iD^\mu H^k H^l$
& $n_f^4$
\\\cline{3-4}
&
& ${\cal O}_{eLeLH^2D2}=\epsilon_{ik}\epsilon_{jl}(\overline{e}L^i)(\overline{e^{\C}}\gamma_\mu L^j) H^k iD^\mu H^l$
& $n_f^4$
\\\cline{3-4}
&
& ${\cal O}_{eeLLH^2D}=\epsilon_{ik}\epsilon_{jl}(\overline{e}\gamma_\mu e)(\overline{L^{\C,i}}i\overleftrightarrow{D}^\mu L^j)H^k H^l$
& ${1\over 2}n_f^3(n_f-1)$
\\\cline{2-4}
&\multirow{3}{*}{$L^3L^*H^2D$}
& ${\cal O}_{LLLLH^2D1}=\delta_{ik}\epsilon_{jm}\epsilon_{ln}(\overline{L^i}\gamma_\mu L^j)(\overline{L^{\C,k}} L^l) i D^\mu H^m H^n$
& $n_f^4$
\\\cline{3-4}
&
& ${\cal O}_{LLLLH^2D2}=\delta_{ik}\epsilon_{jl}\epsilon_{mn}(\overline{L^i}\gamma_\mu L^j)(\overline{L^{\C,k}} L^l) i D^\mu H^m H^n$
& ${1\over 3}n_f^2(2n_f^2+1)$
\\\cline{3-4}
&
& ${\cal O}_{LLLLH^2D3}=\epsilon_{ik}\epsilon_{jl}(\overline{L}\gamma_\mu L)(\overline{L^{\C,i}}  i\overleftrightarrow{D}^\mu L^j)H^k H^l$
& ${1\over 2}n_f^3(n_f-1)$
\\
\hline
\end{tabular}
}
\caption{Dim-9 operators in class $D\psi^4\varphi^2$ for sector $(\Delta B,\Delta L)=(0,2)$. There are 35 operators without counting fermion generations and 30 remain for one generation case. Hermitian conjugated operators are not included or counted.}
\label{tab:LV3}
\end{table}

\begin{table}[H]
\centering
\resizebox{\linewidth}{!}{
\renewcommand{\arraystretch}{1.25}
\begin{tabular}{|l |c| l |c|}
\hline
Classes & Configurations & Operators &\# with $ n_f$
\\\hline\hline
\multicolumn{4}{|c|}{{\color{magenta}$(\Delta B, \Delta L)=(0,2)$}}
\\\hline
\multirow{12}{*}{$D\psi^4X$}
& \multirow{4}{*}{$ud^*L^2 BD$}
& ${\cal O}_{dLLuBD1}=g_1\epsilon_{ij}(\overline{d}D_\mu L^i)(\overline{ L^{\C, j}}\gamma_\nu u)B^{\mu\nu}$
& $n_f^4$
\\\cline{3-4}
&
& ${\cal O}_{dLuLBD2}=g_1\epsilon_{ij}(\overline{d} L^i)(\overline{D_\mu L^{\C, j}}\gamma_\nu u)B^{\mu\nu}$
& $n_f^4$
\\\cline{3-4}
&
&${\cal O}_{dLuLBD3}=g_1\epsilon_{ij}(\overline{d} L^i)(\overline{ L^{\C, j}}\gamma_\mu D_\nu u)B^{\mu\nu}$
& $n_f^4$
\\\cline{3-4}
&
& ${\cal O}_{duLLBD}=g_1\epsilon_{ij}(\overline{d}\gamma_\mu u)(\overline{ L^{\C, i}}\overleftrightarrow {D_\nu} L^j)\tilde B^{\mu\nu}$
& ${1\over 2}n_f^3(n_f+1)$
\\\cline{2-4}
& \multirow{4}{*}{$ud^*L^2 WD$}
& ${\cal O}_{dLLuWD1}=g_2(\epsilon W^{\mu\nu})_{ij}(\overline{d}D_\mu L^i)(\overline{ L^{\C, j}}\gamma_\nu u)$
& $n_f^4$
\\\cline{3-4}
&
& ${\cal O}_{dLuLWD2}=g_2(\epsilon W^{\mu\nu})_{ij}(\overline{d} L^i)(\overline{D_\mu L^{\C, j}}\gamma_\nu u)$
& $n_f^4$
\\
\hline
\end{tabular}
}
\caption{Dim-9 operators in classes $D\psi^4X,~D^2\psi^2\varphi^4, ~D^2\psi^2\varphi^2X$ for sector $(\Delta B,\Delta L)=(0,2)$. There are 34 operators without counting fermion generations and 30 remain for one generation case. Hermitian conjugated operators are not included or counted.}
\label{tab:LV4}
\end{table}

\begin{table}[H]
\centering
\resizebox{\linewidth}{!}{
\renewcommand{\arraystretch}{1.15}
\begin{tabular}{|l |c| l |c|} 
\hline
Classes & Configurations & Operators &\# with $ n_f$
\\\hline\hline
\multicolumn{4}{|c|}{{\color{magenta}$(\Delta B, \Delta L)=(0,2)$}}
\\
\hline
\multirow{28}{*}{$D^2\psi^4\varphi$}
& \multirow{4}{*}{$L^3e^*HD^2$}
& ${\cal O}_{eLLLH^21}=\epsilon_{ij}\epsilon_{kl}(\overline{e}D_\mu L^i)(\overline{D^\mu L^{\C,j}} L^k)H^l$
& $n_f^4$
\\\cline{3-4}
&
& ${\cal O}_{eLLLH^22}=\epsilon_{ij}\epsilon_{kl}(\overline{e}D_\mu L^i)(\overline{L^{\C,j}}D^\mu  L^k)H^l$
& $n_f^4$
\\\cline{3-4}
&
& ${\cal O}_{eLLLH^23}=\epsilon_{ij}\epsilon_{kl}(\overline{e}L^i)(\overline{ L^{\C,j}}D^\mu L^k)D_\mu H^l$
& $n_f^4$
\\\cline{3-4}
&
& ${\cal O}_{eLLLH^24}=\epsilon_{ij}\epsilon_{kl}(\overline{e}L^i)(\overline{D_\mu  L^{\C,j}}L^k)D^\mu H^l$
& ${1\over 3}n_f^2(n_f^2-1)$
\\\clineB{2-4}{3}
& \multirow{7}{*}{$ud^*LeHD^2$}
&  ${\cal O}_{dLeuH^21}=\epsilon_{ij}(\overline{d}D_\mu L^{i})(\overline{D^\mu e^\C}u)H^j$
& $n_f^4$
\\\cline{3-4}
&
& ${\cal O}_{dLeuH^22}=\epsilon_{ij}(\overline{d}D_\mu L^{i})(\overline{e^\C}D^\mu u)H^j$
& $n_f^4$
\\
\hline
\end{tabular}
}
\caption{Dim-9 operators in classes $D^2\psi^4\varphi,~D^3\psi^2\varphi^3,~D^4\psi^4,~D^4\psi^2\varphi^2$ for sector $(\Delta B,\Delta L)=(0,2)$. There are 35 operators without counting fermion generations and 34 remain for one generation case. Hermitian conjugated operators are not included or counted.}
\label{tab:LV5}
\end{table}

\begin{table}[H]
\centering
\resizebox{\linewidth}{!}{
\renewcommand{\arraystretch}{1.32}
\begin{tabular}{|l |c| l |c| }
\hline
Classes & Configurations & Operators &\# with $ n_f$
\\\hline\hline
\multicolumn{4}{|c|}{{\color{magenta}$(\Delta B, \Delta L)=(2,0)$}}
\\
\hline
\multirow{5}{*}{$\psi^6$}
&  \multirow{2}{*}{$d^2Q^4$}
&   ${\cal O}_{ddQQQQ1}=\epsilon_{\alpha\beta\gamma}\epsilon_{\rho\sigma\tau}\epsilon_{ik}\epsilon_{jl}(\overline{d^{\C,\alpha}} d^\rho)(\overline{Q^{\C,i\beta}} Q^{j\sigma})(\overline{Q^{\C,k\gamma}} Q^{l\tau})$
&\multirow{2}{*}{${1\over 12}n_f^3(n_f+1)(5n_f^2-6n_f+7)$}
\\
&
&   ${\cal O}_{ddQQQQ2}=\epsilon_{\alpha\beta\tau}\epsilon_{\rho\sigma\gamma}\epsilon_{ik}\epsilon_{jl}(\overline{d^{\C,\alpha}} d^\rho)(\overline{Q^{\C,i\beta}} Q^{j\sigma})(\overline{Q^{\C,k\gamma}} Q^{l\tau})$
&
\\
\cline{2-4}
& $d^3Q^2u$
&  ${\cal O}_{udddQQ}=\epsilon_{\alpha\beta\gamma}\epsilon_{\rho\sigma\tau}\epsilon_{ij}(\overline{u^{\C,\alpha}} d^\rho)(\overline{d^{\C,\beta}}d^\sigma)(\overline{Q^{\C,i\gamma} } Q^{j,\tau})$
& ${1\over 6}n_f^3(5n_f^3+n_f^2+n_f-1)$
\\
\cline{2-4}
&  \multirow{2}{*}{$d^4u^2$ }
&  ${\cal O}_{ududdd1}=\epsilon_{\alpha\beta\gamma}\epsilon_{\rho\sigma\tau}(\overline{u^{\C,\alpha}}d^\rho)(\overline{u^{\C,\beta}}d^\sigma)(\overline{d^{\C,\gamma}}d^\tau) $
&\multirow{2}{*}{${1\over 48}n_f^2(n_f+1)(25n_f^3-18n_f^2+35n_f+6)$}
\\
&
&  ${\cal O}_{ududdd2}=\epsilon_{\alpha\sigma\gamma}\epsilon_{\rho\beta\tau}(\overline{u^{\C,\alpha}}d^\rho)(\overline{u^{\C,\beta}}d^\sigma)(\overline{d^{\C,\gamma}}d^\tau) $
&
\\
\hline\hline
\multicolumn{4}{|c|}{{\color{magenta}$(\Delta B, \Delta L)=(1,3)$}}
\\
\hline
\end{tabular}
}
\caption{Dim-9 operators in sectors $(\Delta B,\Delta L)=(2,0),(1,3)$, and in classes $\psi^4\varphi X,~\psi^4\varphi^3$ for sector $(\Delta B,\Delta L)=(1,-1)$. There are 40 operators without counting fermion generations and 29 remain for one generation case. Hermitian conjugated operators are not included or counted.}
\label{tab:BV1}
\end{table}

\begin{table}[H]
\centering
\resizebox{\linewidth}{!}{
\renewcommand{\arraystretch}{1.13}
\begin{tabular}{|l |c| l |c|}
\hline
Classes & Configurations & Operators &\# with $ n_f$
\\\hline\hline
\multicolumn{4}{|c|}{{\color{magenta}$(\Delta B, \Delta L)=(1,-1)$}}
\\
\hline
\multirow{24}{*}{$\psi^6$}
& $d^3ee^{*2}$
& ${\cal O}_{eeeddd}=\epsilon_{\alpha\beta\gamma}(\overline{e}e^\C)(\overline{e^\C}d^\alpha)(\overline{d^{\C,\beta}} d^\gamma)$
& ${1\over 6}n_f^3(n_f+1)(n_f^2-1)$
\\\cline{2-4}
& $d^3LL^*e^* $
& ${\cal O}_{eLLddd}=\epsilon_{\alpha\beta\gamma}(\overline{e}L)(\overline{L}d^\alpha)(\overline{d^{\C,\beta}} d^\gamma)$
& ${1\over 3}n_f^4(n_f^2-1)$
\\\cline{2-4}
& \multirow{2}{*}{$d^2ueL^{*2} $}
& ${\cal O}_{LeLdud}=\epsilon_{\alpha\beta\gamma}\epsilon_{ij}(\overline{L^i}e)(\overline{L^j} d^\alpha)(\overline{u^{\C,\beta}} d^\gamma)$
& $n_f^6$
\\\cline{3-4}
&
& ${\cal O}_{LLeudd}=\epsilon_{\alpha\beta\gamma}\epsilon_{ij}(\overline{L^i}\sigma_{\mu\nu}L^{\C,j})(\overline{e^\C}u^\alpha)(\overline{d^{\C,\beta}}\sigma^{\mu\nu}d^\gamma)$
& ${1\over 4}n_f^4(n_f+1)^2$
\\\cline{2-4}
& \multirow{1}{*}{$Qd^2eL^*e^* $}
& ${\cal O}_{eQLded}=\epsilon_{\alpha\beta\gamma}(\overline{e}Q^\alpha)(\overline{L}d^\beta)(\overline{e^\C}d^\gamma)$
& $n_f^6$
\\\cline{2-4}
& \multirow{1}{*}{$Qd^2LL^{*2}$}
& ${\cal O}_{LQLdLd}=\epsilon_{\alpha\beta\gamma}\delta_{ik}\delta_{jl}(\overline{L^{\C,i}} Q^{j\alpha})(\overline{L^k} d^\beta)(\overline{L^l} d^\gamma)$
& $n_f^6$
\\
\hline
\end{tabular}
}
\caption{Dim-9 operators in classes $\psi^6,~D\psi^4\varphi^2$ for sector $(\Delta B,\Delta L)=(1,-1)$. There are 44 operators without counting fermion generations and 39 remain for one generation case. The $SU(3)_C$ contraction with $\delta_{\alpha\beta}$ is implied in class $\psi^6$ class. Hermitian conjugated operators are not included or counted.}
\label{tab:BV2}
\end{table}

\begin{table}[H]
\centering
\resizebox{\linewidth}{!}{
\renewcommand{\arraystretch}{1.13}
\begin{tabular}{|l |c| l |c|}
\hline
Classes & Configurations & Operators &\# with $ n_f$
\\\hline\hline
\multicolumn{4}{|c|}{{\color{magenta}$(\Delta B, \Delta L)=(1,-1)$}}
\\\hline
\multirow{20}{*}{$D\psi^4X$}
& \multirow{2}{*}{$d^3e^* BD$ }
& ${\cal O}_{edddBD1}=g_1\epsilon_{\alpha\beta\gamma}(\overline{e}\gamma_{\mu}d^\alpha)(\overline{d^{\C,\beta}}D_\nu d^\gamma)B^{\mu\nu}$
&  $n_f^4$
\\\cline{3-4}
&
& ${\cal O}_{edddBD2}=g_1\epsilon_{\alpha\beta\gamma}(\overline{e}\gamma_{\mu}d^\alpha)(\overline{d^{\C,\beta}}D_\nu d^\gamma)\tilde B^{\mu\nu}$
& ${1\over 6}n_f^2(n_f^2+3n_f+2)$
\\\cline{2-4}
& \multirow{3}{*}{$d^3e^* GD$ }
& ${\cal O}_{edddGD1}=g_3(\epsilon G^{\mu\nu})_{\alpha\beta\gamma}(\overline{e}\gamma_{\mu}d^\alpha)(\overline{D_\nu d^{\C,\beta}}d^\gamma)$
& $n_f^4$
\\\cline{3-4}
&
& ${\cal O}_{edddGD2}=g_3(\epsilon G^{\mu\nu})_{\alpha\beta\gamma}(\overline{e}\gamma_{\mu}d^\alpha)(\overline{ d^{\C,\beta}}D_\nu d^\gamma)$
&  $n_f^4$
\\\cline{3-4}
&
& ${\cal O}_{edddGD3}=g_3(\epsilon \tilde G^{\mu\nu})_{\alpha\beta\gamma}(\overline{e}\gamma_{\mu}d^\alpha)(\overline{d^{\C,\beta}}D_\nu d^\gamma)$
& ${1\over 3}n_f^2(n_f^2-1)$
\\\cline{2-4}
& \multirow{4}{*}{$Qd^2L^* BD$ }
&  ${\cal O}_{LddQBD1}=g_1\epsilon_{\alpha\beta\gamma}(\overline{L}D_\mu d^\alpha)(\overline{d^{\C,\beta}}\gamma_\nu Q^\gamma)B^{\mu\nu}$
& $n_f^4$
\\
\hline
\end{tabular}
}
\caption{Dim-9 operators in classes $D\psi^4X,~D^2\psi^4\varphi,~D^3\psi^4$ for sector $(\Delta B,\Delta L)=(1,-1)$. There are 41 operators without counting fermion generations and 38 remain for one generation case. Hermitian conjugated operators are not included or counted.}
\label{tab:BV3}
\end{table}


\bibliographystyle{JHEP}
\bibliography{refs}

\end{document}